\documentclass[journal,twoside]{IEEEtran}
\usepackage{cite}
\usepackage{amsmath,amssymb,amsfonts}
\usepackage{algorithmic}
\usepackage{graphicx}
\usepackage{textcomp}
\usepackage{subfigure}
\usepackage{booktabs}
\usepackage{makecell}
\usepackage{multirow}

\def\BibTeX{{\rm B\kern-.05em{\sc i\kern-.025em b}\kern-.08em
    T\kern-.1667em\lower.7ex\hbox{E}\kern-.125emX}}
\begin{document}

\title{Transfer Deep Reinforcement Learning-based Large-scale V2G Continuous Charging Coordination with Renewable Energy Sources}
\author{Yubao~Zhang, Xin~Chen* and Yuchen~Zhang
   
 \thanks{This work was supported in part by the National Natural Science Foundation of China (Grant No.21773182 (B030103)) and the HPC Platform, Xi'an Jiaotong University. (Corresponding author: Xin Chen, e-mail: xin.chen.nj@xjtu.edu.cn)}
 \thanks{Yubao~Zhang is with Center of Nanomaterials for Renewable Energy, State Key Laboratory of Electrical Insulation and Power Equipment, School of Electrical Engineering, Xi'an Jiaotong University, Xi'an, Shaanxi, China (e-mail: YubaoZhang@stu.xjtu.edu.cn).}
\thanks{Xin Chen is with Center of Nanomaterials for Renewable Energy, State Key Laboratory of Electrical Insulation and Power Equipment,  School of Electrical Engineering, Xi'an Jiaotong University, Xi'an, Shaanxi, China (e-mail: xin.chen.nj@xjtu.edu.cn). }
\thanks{Yuchen Zhang is with School of Electrical Information, Xi'an Jiaotong University, Xi'an, Shaanxi, China (e-mail: zyc20000201@stu.xjtu.edu.cn). }
}
\maketitle

\begin{abstract}
Due to the increasing popularity of electric vehicles (EVs) and the technological advancement of EV electronics, the vehicle-to-grid (V2G) technique and large-scale scheduling algorithms have been developed to achieve a high level of renewable energy and power grid stability. 
This paper proposes a deep reinforcement learning (DRL) method for the continuous charging/discharging coordination strategy in aggregating large-scale EVs in V2G mode with renewable energy sources (RES).
The DRL coordination strategy can efficiently optimize the electric vehicle aggregator's (EVA's) real-time charging/discharging power with the state of charge (SOC) constraints of the EVA and the individual EV. Compared with uncontrolled charging, the load variance is reduced by 97.37$\%$ and the charging cost by 76.56$\%$. The DRL coordination strategy further demonstrates outstanding transfer learning ability to microgrids with RES and large-scale EVA, as well as the complicated weekly scheduling. The DRL coordination strategy demonstrates flexible, adaptable, and scalable performance for the large-scale V2G under realistic operating conditions.  
\end{abstract}


\section{Introduction}
\label{sec:introduction}
\subsection{Background and Motivation}
\IEEEPARstart{R}{ecent} advances in battery storage technologies, that lowered the battery prices, together with unprecedented awareness towards $CO_2$ emissions, created momentum for electrification in the transportation sector \cite{tookanlou2021optimal}. Against this background, vehicle-to-grid (V2G)  technology can promote the growth and development of electric transportation with a low carbon footprint and cost saving for consumers  \cite{singh2020cost}. However,  significant concerns and challenges for power systems operations  are posed due to uncertainties in electric vehicle (EV) users' charging behavior.

\paragraph{EV/V2G Coordination Strategy}

EVs can effectively promote the efficient utilization of energy and decrease the power grid load fluctuation, which provides a broad prospect for the efficient, flexible and large-scale application of distributed renewable energy sources (RES). The orderly scheduling strategy of EVs/V2G has attracted the attention of researchers. Dabbaghjamanesh et al. \cite{dabbaghjamanesh2020stochastic} proposed a newly stochastic framework based on unscented transform to model uncertainties in EVs, e.g., the charging strategy and number of EVs to be charged. To model the competitive market for EV charging service providers, a versatile game-theoretic model was proposed by which each EV charging service provider aims to maximize individual profit through optimizing locations, numbers, the price of the service \cite{zavvos2019planning}. As for the charging scheduling strategy, in \cite{zhu2018coordinated}, a coordinated sectional droop charging control strategy was proposed for an electric vehicle aggregator (EVA) that participates in the frequency regulation of the microgrids with high penetration RES. In \cite{li2021data}, a deep learning technique involving a multi-relation graph convolutional network was created to forecast the EV charging demand. In \cite{peng2017optimal}, an optimal dispatching strategy for V2G aggregator was proposed to satisfy the driving demand of EV owners and maximize the economic benefits of aggregator simultaneously. 

In order to investigate the impact of EVs' charging/discharging behavior and demand side response resources on the economic operation of photovoltaic (PV) grid-connected microgrid system, a multi-objective model of microgrid economic scheduling was proposed \cite{hou2020multi}. To reduce EVA operation costs and maximize the travel utility for EV users participating in this service, Jin et al. developed an EVA optimization schedule model that combined a day-ahead optimization schedule and real-time optimization schedule \cite{jin2020optimal}. The negative impact of large-scale PV and EV integration
on distribution network is mutually reduced via V2G technology\cite{mouli2017integrated}.

Swarm intelligent algorithms are widely used in EV scheduling problems, such as particle swarm optimization (PSO) \cite{badawy2016power}, genetic algorithm \cite{mao2020electric} and so on. Because swarm intelligent  algorithms are less dependent on mathematical model compared to classical analytical methods of optimization \cite{li2019nash}. However, most swarm intelligence algorithms fall into local optimum easily, and convergence speed is very slow. Besides, when the optimization objective changes, the algorithm needs to be retrained and calibrated for new parameters.

Reinforcement learning (RL) introduces ambient intelligence into the systems by providing a class of solution methods to the closed loop problem of processing the data to generate control decisions to react\cite{lei2020deep}. Specifically, the agents interact with the environment to learn optimal policies that map status or states to actions\cite{sutton2011reinforcement}. 

\paragraph{Deep Reinforcement Learning}
Deep reinforcement learning (DRL) is characterized by a combination of RL and deep learning algorithms. Deep learning analyzes environmental information and extracts features from it; RL analyzes new environmental features based on these features, and have the ability to select actions in the new environment to achieve target rewards.  

At present, many DRL algorithms have been proposed and applied to power system research, including the use of deep Q network (DQN)\cite{huang2019adaptive} and double DQN (to avoiding the problem of overestimation of the $Q$ value in the DQN algorithm)\cite{yang2019two} to optimize the voltage control of the distribution network; The dueling deep Q network\cite{wang2016dueling} was used to solve the demand-side response problem\cite{wang2020deep}, and it improved the DQN algorithm network structure\cite{van2016deep}, which can solve the problem of overestimation of the DQN value function, thereby enhancing the generalization ability of the model and improving the stability of the model. DQN can learn successful policies directly from sensory inputs using end-to-end DRL \cite{mnih2015human}. However, DQN can only handle discrete and low-dimensional action spaces\cite{lillicrap2015continuous}. Trust Region Policy Optimization (TRPO) is effective and robust for optimizing large nonlinear policies such as neural networks\cite{schulman2015trust}. Despite its approximations that deviate from the theory, TRPO tends to give monotonic improvement, with little tuning of hyperparameters. Proximal policy optimization (PPO), has some of the benefits of TRPO, but is much simpler to implement, more general, and have better sample complexity (empirically). Besides overall strikes a favorable balance between sample complexity, simplicity, and wall-time \cite{schulman2017proximal}. Based on the fluctuations in the output of RES, the DRL optimization study of EV charging control was carried out\cite{chen2021electric}, but the charging and discharging characteristics of the EV itself were not discussed in depth. In addition, for an individual EV, a DRL real-time scheduling method that considers the randomness of EV user behavior and the uncertainty of real-time electricity prices was proposed \cite{wan2018model}. Yan et al. aimed to crack the individual EV charging scheduling problem based on DRL considering the driver's anxiety and the charging cost \cite{yan2021deep}. To address this challenge on the online scheduling of shared autonomous EV fleets, Qian et al. proposed a novel framework named the shadow-price DRL\cite{qian2022shadow}. To dynamically adjust the EV price in a timely fashion to unlock the flexibility of EV customers, high-dimensional continuous charging cases were studied in \cite{zhao2021dynamic, qiu2020deep}. The DRL algorithms use reward and penalty functions to characterize user travel needs, and don't rely on physical models.
It is noticed that traditional swarm intelligent algorithms still prevail in large-scale V2G charging/discharging coordination problems. However, the optimization process of these algorithms are costly, which can not meet the needs of large-scale V2G real-time scheduling.

\subsection{Scope and Contributions}
This article proposes a novel DRL coordination strategy for peak cutting and valley filling to ensure the safe and stable operation of the power grid, accounting for the large-scale V2G continuous charging/discharging power. In order to schedule large-scale V2G efficiently, the DRL coordination strategy should take into consideration the EVA state of charge (SOC) constraints and EVA power allocation to individual EV. Then, the PPO algorithm is adopted to coordinate the large-scale V2G continuous charging/discharging problems. In contrast with traditional swarm intelligent algorithms and previous DRL algorithms, the proposed strategy optimized by the PPO algorithm sets up the difficulty in multi-dimensional state and continuous action spaces. Experiment results show that the proposed strategy is effective in suppressing grid load fluctuations and can be transferred to RES scenario, large-scale EVA scenario and weekly scenario. More specifically, our contributions of this article are outlined in the following.

First, the DRL formulation is proposed for the large-scale V2G continuous charging/discharging coordination to reduce the load variance in the consideration of the EVA state of charge (SOC) constraints and EVA power allocation to individual EV.

Second, The DRL coordination strategy can theoretically realize real-time precise continuous adjustment of charging/discharging power, because the PPO algorithm can solve the problems in multi-dimensional state and continuous action spaces, in contrast with discrete DRL algorithms such as DQN. 

Finally, experiment results demonstrate that the proposed strategy achieves a significantly better peak shaving and valley filling performance in contrast with the PSO and DQN algorithms. Compared with uncontrolled charging, the proposed strategy reduces load variance by 97.37$\%$ and charging cost by 76.56$\%$. Besides, the DRL coordination strategy further demonstrates outstanding transfer ability to microgrids with RES and different sizes of EVA, as well as the complicated long-time scale weekly scheduling.

\section{Coordination Problem Formulation}

The DRL coordination test system is composed of wind turbines (WTs), PVs, and EVA, and these distributed energy installations are collectively managed by a central control entity with the PPO algorithm. It is reported that EVs will be parked at home for a majority of the time, and the batteries only require to be charged with enough energy before their departure time. Hence, EVs can collectively serve as large-capacity energy storage to mitigate the volatility and intermittency of PV and WT outputs, and facilitate the scheduling system to boost its total profit by moving electricity delivery through hours. 

An EVA represents numerous EVs in the wholesale market and coordinates their operation according to the power load conditions and the EV operating characteristics such as traveling patterns and battery's/charger's operating parameters to optimize the corresponding objective functions. By only revealing the aggregate charging/discharging demand, the operators do not need to disclose all of their customers' charging/discharging information to municipal operators and therefore protect the customer's privacy. In addition, with only the aggregate charging/discharging requirement information of EVA, the municipal operator can optimize the charging/discharging power trajectory for the EVA with considering the detailed charging/discharging needs of individual EV customers.

An EVA agent is a technical and commercial bridge between a system operator and EVs. Optimal participation of an EVA in day-ahead energy and reserve markets demands a DRL coordination strategy. From the system operator perspective, the aggregator is seen as a large source of generation or load, which provides ancillary services such as spinning and regulating reserve. With the DRL coordinated strategy at the EVA level, power grid can reduce the power load fluctuation.

The real-time SOC of the EVA is critical to the measurement of the real-time state of EVA and the EV charging/discharging management. 
The EVA SOC is enforced to meet the SOC travel demand of users. The EVA SOC  at the $k$-th timeslot, $SOC^{k}$, is defined as, 
\begin{equation}\label{ssoc}
SOC^{k}=\frac{P_{EVA}^{k}\Delta t + \sum\limits_{n=1}^{N}{SOC_{n}^{k}}L_{n}^{k}Q_n}{\sum\limits_{n=1}^{N}L_{n}^{k}Q_n},
\end{equation}
where
\begin{align}
\begin{split}
&L_{n}^{k}= \left \{ 
\begin{array}{cc} 
    1, &arrival\\ 
   0,  & departure 
\end{array}
\right.
\end{split}.
\end{align}

$SOC_{n}^{k}$ is the SOC of the $n-th$ EV at the $k-th$ timeslot, $P_{EVA}^k$ is the charging/discharging power of EVA at the $k$-th timeslot, $Q_n$ is the battery capacity of the $n-th$ EV. $\Delta t$ is the length of the timeslot, which is one hour. $L_{n}^{k}$ is the scheduling state function of the $n-th$ EV at the timeslot $k$. When $L_{n}^{k}=1$, it means that the EV arrives home ready for the charging/discharging scheduling at the timeslot $k$. When $L_{n}^{k}=0$, it means that the EV leaves home and can't be scheduled at the timeslot $k$.

The DRL coordination strategy optimizes the battery states of the examined EVA as an intermediary entity between the EVs and the power grid. The problem is formulated as follows:
\begin{equation}
\max\limits_{P_{EVA}^{k}}{\frac{\alpha}{\sigma^{2}}}+\beta\left(P_{\max }-P_{\min }\right)+\chi\left\lvert SOC_{\text {b}}^{k}-SOC^k\right\rvert,\label{goal}
\end{equation}
subject to:
\begin{align}
&{SOC_{min}^{k} \leq SOC^{k}\leq SOC_{max}^{k}},\label{soccon}\\
&P_{EVA,dis}^{max} \leq P_{EVA}^{k} \leq P_{EVA,ch}^{max},\label{peva}\\
&P_{grid}^{min} \leq P^k \leq P_{grid}^{max},\label{ptie}
\end{align}
where,
\begin{align}
&P_{\max }=\max{(P^{k-23},...,P^k)},\label{maxpower}\\
&P_{\min }=\min{(P^{k-23},...,P^k)},\label{minpower}\\
&P^k=P_{base}^k+P_{EVA}^{k},\label{powerload}\\
&\sigma^{2}=\frac{1}{K} \sum_{k=1}^{K}\left(P^k-\bar{P}\right)^{2},\label{variance}\\
&\bar{P}=\frac{1}{K} \sum_{k=1}^{K}\left(P^k\right),\label{ap}\\
&C= \sum_{k=1}^K\sum_{n=1}^{N}c^{k}p_{n}^{k}L_{n}^{k}\Delta t.\label{costm}
\end{align}

The objective function (\ref{goal}) maximizes the overall profit of the EVA and the power grid, which includes the following components: 1) the power grid load variance ${\sigma^{2}}$; 2) its difference between peak load $P_{\max }$ and valley load $P_{\min}$; and 3) the EVA SOC . And the solution variable is ${P_{EVA}^{k}}$. $\alpha$ is the reward coefficient of load variance, which is positive; $\beta$ is the penalty coefficient of peak-valley difference of power grid, which is negative; $\chi$ is a positive constraint coefficient of EVA SOC, which ensures that the SOC does not exceed the boundary value; $SOC_{\text {b}}^{k}$ is the SOC boundary value of EVA at the timeslot $k$, which includes $SOC_{min}^k$ and $SOC_{max}^k$.

The EVA constraints are shown as (\ref{soccon})-(\ref{ptie}). The SOC constraint is described in (\ref{soccon}), where $SOC_{min}^k$ and $SOC_{max}^k$ are the minimum and maximum SOC of the EVA at the timeslot $k$ respectively. 
The power of EVA constraint is described in (\ref{peva}), where $P_{EVA,dis}^{max}$ and $P_{EVA,ch}^{max}$ are the maximum discharging power and maximum charging power of the EVA respectively. 
The tie-line power constraint is described in (\ref{ptie}), where $P^{k}$ is the power load at the timeslot $k$ calculated as (\ref{powerload}), $P_{grid}^{min}$ and $P_{grid}^{max}$ are the minimum and maximum tie-line power respectively. 

$P_{\max }$ and $P_{\min }$ are the maximum and minimum value of power load respectively after scheduling, the calculation is shown in (\ref{maxpower}) and (\ref{minpower}).
The load variance $\sigma^{2}$ is an indicator reflecting the stability of power grid load. The smaller its value is, the more stable the grid load is. The mathematical model of the load variance is shown as (\ref{variance}), $K$ is the number of timeslots. 
$\bar{P}$ is the average of the total power load during the day, which is shown as (\ref{ap}), $P_{base}^k$ is the baseload of at the timeslot $k$. 
$C$ is the EV's charging/discharging cost or profit, the positive value is cost and the negative value is the profit. The mathematical model of the cost $C$ is shown as (\ref{costm}). $p_{n}^{k}$ is the charging/discharging power of the $n-th$ EV at the timeslot k. $c^{k}$ is the time of use (TOU) tariff at the timeslot $k$. 

Based on the EVA charging/discharging scheduling, individual EV SOC constraints need to be satisfied. The consensus algorithm has the advantages of convenience, speed and practicality in solving multi-agent co-optimization problems, especially power dynamic allocation problems \cite{zhang2016virtual}. Therefore, to obtain the individual EV power, an EVA power allocation algorithm based on SOC buffer consistency is proposed. The SOC buffer factor $\kappa$ can be defined as:

\begin{align}
&\kappa=\left\{\begin{array}{cc}
\frac{(SOC_{n}^{k}-SOC_{n}^{k-1})}{{SOC_{n,max}^{k}}-SOC_{n}^{k-1}} &P_{EVA}^{k}>0 \\
\frac{(SOC_{n}^{k-1}-SOC_{n}^{k})}{SOC_{n}^{k-1}-{SOC_{n,min}^{k}}} &P_{EVA}^{k}<0 
\end{array}\right.,\label{kappa}\\
&p_{n}^{k^{*}}=(SOC_{n}^{k}-SOC_{n}^{k-1})*{Q_n}/\Delta t,\\
&P_{EVA}^{k}={\sum\limits_{n=1}^{N}{p_{n}^{k^{*}}L_{n}^{k}}},\label{powerall}
\end{align}
subject to
\begin{align}
&SOC_{n,min}^{k}\leq SOC_{n}^{k}\leq SOC_{n,max}^{k}.\label{deps}
\end{align}

Where $SOC_{n,min}^{k}$ and $SOC_{n,max}^{k}$ are the minimum and maximum SOC for the $n-th$ EV at the timeslot $k$ respectively. In particular, when ${SOC_{n,max}^{k}}=SOC_{n}^{k-1}$ and $P_{EVA}^{k}>0$, $p_{n}^{k^{*}}=0$; when ${SOC_{n,min}^{k}}=SOC_{n}^{k-1}$ and $P_{EVA}^{k}<0$, $p_{n}^{k^{*}}=0$.

After calculating $p_{n}^{k^{*}}$, a safety check correction is needed to obtain the true charging/discharging power $p_{n}^{k}$. During scheduling, the upper and lower limits of the output power of each EV are used to correct the EV SOC, because the maximum charging and discharging power should not be exceeded. An individual EV should meet the power constraint, the charging/discharging power $p_{n}^{k}$ is corrected as below.

\begin{align}\label{soc1}
&p_{n}^{k}= \begin{cases}p_{n}^{k^{*}} & p_{dis}^{max} \leqslant p_{n}^{k^{*}} \leqslant p_{ch}^{max} \\
 p_{ch}^{max} & P_{EVA}^{k}>0, p_{n}^{k^{*}}>p_{ch}^{max} \\
 p_{dis}^{max} & P_{EVA}^{k}<0, p_{n}^{k^{*}}<p_{dis}^{max}\end{cases}.
\end{align}

Where $p_{ch}^{max}$ and $p_{dis}^{max}$ are the maximum charging and discharging power of the EV, respectively. It's worth mentioning that the SOC buffer factor $\kappa$ during the allocation of each timeslot, and its value is jointly determined by each EV SOC, the EVA power and the EVA SOC. To allocate the EVA power to the individual EV correctly, all the EVs should have the same SOC buffer factor $\kappa$ in a timeslot. Therefore the power allocation, $P_{EVA}^{k}$, is carried out when all the EVs have the same buffer factor $\kappa$.
During scheduling, EVs with larger SOCs take a larger share of power when $P_{EVA}^{k}>0$; EVs with smaller SOCs take a larger share of power when $P_{EVA}^{k}<0$. The SOC of each EV can effectively converge and all reach the desired SOC before their departure time, solving the problem of the redistribution caused by the initial SOC difference.

The V2G technology can improve the penetration of intermittent renewable energy in microgrids. The microgrid includes a variety of RES (such as wind power, PV power, etc.). With RES, the goal in (\ref{pwgoal}) of the DRL coordination strategy is to increase the consumption of renewable energy as much as possible while meeting the SOC driving demands and reducing the load variance.
\begin{equation}
\max\limits_{P_{EVA}^{k}}{\frac{\alpha}{\sigma^{2}}+\beta\left(P_{\max }-P_{\min }\right)+\chi\left\lvert  SOC_{\text {b}}^{k}-SOC^k\right\rvert}+\frac\psi{f_1},\label{pwgoal}\\
\end{equation}
where
\begin{align}
&f_1 = \frac1K \sum_{k=1}^{K}P_{load}^k,\label{f1}\\
&P_{load}^k = -P_{pv}^k-P_{wt}^k+P_{EVA}^{k},\label{pload}\\
&P^k=P_{base}^k-P_{pv}^k-P_{wt}^k+P_{EVA}^{k},
\end{align}
\begin{align}
&P_{pv}^k=P_{N} \frac{I_{r}^k}{I_{N}}\left[1+\alpha_{T}\left(T^k-T_{N}\right)\right],\label{pv}\\
&P_{wt}^{k}= \begin{cases}0 & 0 \leq v^k \leq v_{c i} \text { or } v^k \geq v_{c o} \\ \frac{v^k-v_{c i}}{v_{r}-v_{c i}} P_{w-\text{rate}} & v_{c i}<v^{k} \leq v_{r} \\ P_{w-rate} & v_{r}<v^{k}<v_{c o}\end{cases}.\label{wt}
\end{align}
where $\psi$ is the reward coefficient of the mean net load of the micro-gird $f_1$, which is negative and ensures the absorption of maximum renewable energy generation. 
The goal with RES is minimizing the $f_1$ as shown in (\ref{f1}). In (\ref{pload}), $P_{load}^k$ is net load of microgrid at the timeslot $k$, which is negative. Because the renewable energy generation is negative, the smaller the $P_{load}^k$, the more renewable energy is consumed.

$P_{pv}^k$ is PV power at the timeslot $k$, and the PV output model is as (\ref{pv}). Where $P_{N}$ is the PV output in standard state, $I_{r}^k$ is the actual solar radiation intensity at time $k$, $I_{N}$ is the solar radiation intensity under ideal conditions, $\alpha_{T}$ is the power temperature coefficient of PV panels, $T^k$ is the actual temperature of PV panels, $T_{N}=25^{\circ}C$.

$P_{wt}^k$ is wind power at the timeslot $k$. Relationships between the output wind power $P_{wt}^k$ and WT wind speed $v^k$ can be represented by piecewise functions, as (\ref{wt}). Where $P_{w-\text{rate}}$ is the rated output power of WT, $v_{ci}$ is the cut-in wind speed, $v_{r}$ is the rated wind speed, $v_{co}$ is the cut-out wind speed.

\section{Proposed Methodology}

In the DRL algorithms, the agent interacts with an environment through a sequence of observations, actions and rewards. The goal of the agent is to select actions in a fashion that maximizes cumulative future reward\cite{mnih2015human}. DRL can be defined as a Markov decision process which includes: 1) a state space $\mathcal{S}$; 2) an action space $\mathcal{A}$; 3) a transition dynamics distribution with conditional transition probability $p\left(s_{k+1} \mid s_{k}, a_{k}\right)$, satisfying the Markov property, i.e., $p\left(s_{k+1} \mid s_{k}, a_{k}\right)=p\left(s_{k+1} \mid s_{1}, a_{1}, \ldots, s_{k}, a_{k}\right) ;$ and 4) a reward $r:$ $\mathcal{S} \times \mathcal{A} \rightarrow \mathbb{R}$.

\begin{figure}[h]
    \centering
    \includegraphics[width=0.5\textwidth]{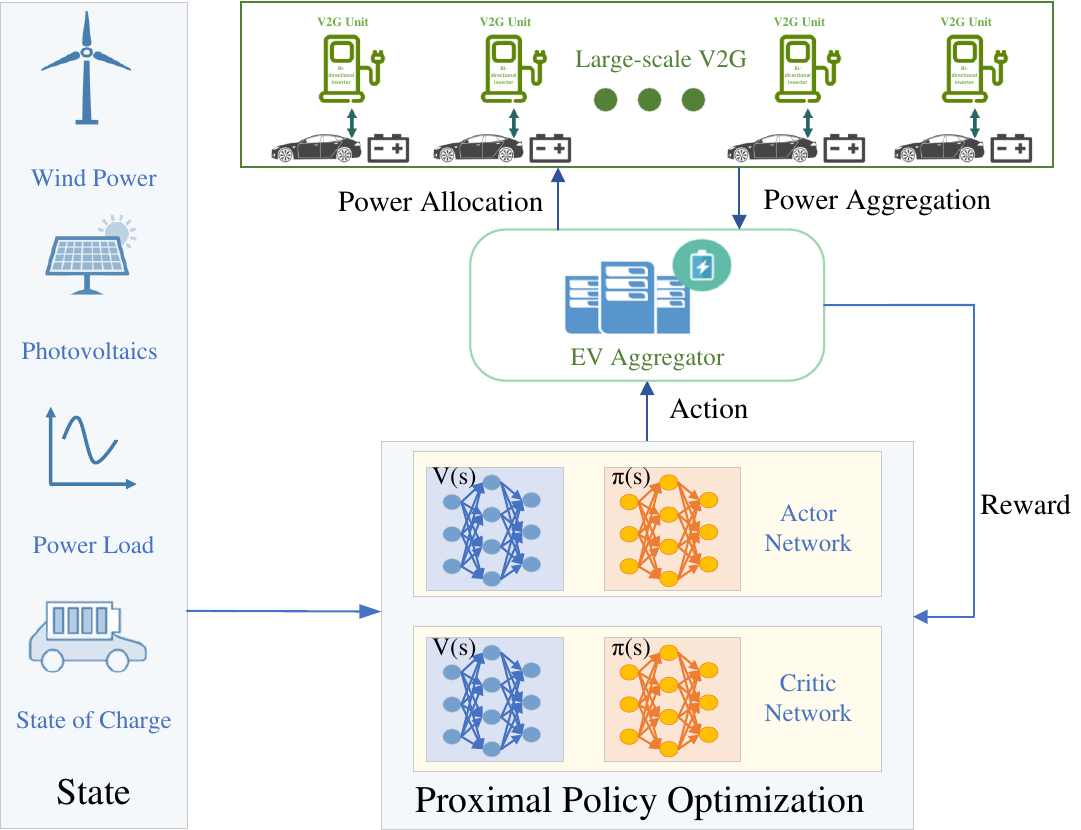}
    \caption{Workflow of proposed DRL coordination strategy by the PPO scheduling.}
    \label{fig:ppoflow}
\end{figure}

PPO, as a class of continuous DRL algorithms, is applied to the DRL coordination strategy for the large-scale EVA that can effectively jointly optimize the grid load. The DRL coordination strategy is presented as Fig.~\ref{fig:ppoflow}.

 At timeslot $k$, we observe the system state $s_k$ which includes the information about the SOC of the EVA, the load variance and the 24-hour load values. Based on this information, the agent will pick the charging/discharging action $a_k$. This action represents the amount of energy that the EVA will be charged or discharged during this timeslot. After executing this action, we can observe the new system state $s_{k+1}$ and choose the new charging/discharging action $a_{k+1}$ for timeslot $k+1$.

\subsection{Proximal Policy Optimization (PPO)}
An actor-critic, model-free algorithm was presented based on the deterministic policy gradient that can operate over continuous action spaces\cite{lillicrap2015continuous}. In both actor network and critic network, there is a value network $V(s)$ and a policy network $\pi(s)$. PPO has some benefits of TRPO, but is much simpler to implement.
PPO algorithm that uses fixed-length trajectory segments is shown below. Each iteration, each of $Y$ (parallel) actors collect $K$ time steps of data. Then we construct the surrogate loss on these $YK$ time steps of data, and optimize it with mini batch stochastic gradient descent (or usually for better performance\cite{kingma2014adam}), for epochs. The main objective of PPO is the following, which is (approximately) maximized each iteration:
\begin{align}
&L_{k}^{CVB}(\theta)=\hat{\mathbb{E}}_{k}\left[L_{k}^{C}(\theta)-c_{1} L_{k}^{VF}(\theta)+c_{2} B\left[\pi_{\theta}\right]\left(s_{k}\right)\right].
\end{align}
Where
\begin{align}
&L_{k}^{C}(\theta)=\hat{\mathbb{E}}_{k}\left[\min \left(L_{k}^{CPI}, G_{k}^{CLIP}\right)\right],\\
&L_{k}^{CPI}=r_{k}(\theta) \hat{A}_{k},\\
&G_{k}^{CLIP}=\operatorname{clip}\left(r_{k}(\theta), 1-\epsilon, 1+\epsilon\right) \hat{A}_{k},\\
&L_{k}^{VF}=\left(V_{\theta}\left(s_{k}\right)-V_{k}^{\operatorname{targ}}\right)^{2},\\
&\hat{A}_{k}=\delta_{k}+(\gamma \lambda) \delta_{k+1}+\cdots+\cdots+(\gamma \lambda)^{K-k+1} \delta_{T-1},\\
&\delta_{k}=r_{k}+\gamma V\left(s_{k+1}\right)-V\left(s_{k}\right),\\
&r_{k}(\theta)=\frac{\pi_{\theta}\left(a_{k} \mid s_{k}\right)}{\pi_{\theta_{\text {old }}}\left(a_{k} \mid s_{k}\right)},
 \text { so } r\left(\theta_{\text {old }}\right)=1.
\end{align}
Where $c_{1}, c_{2}$ are coefficients, and $B$ denotes an entropy bonus, and $L_{k}^{VF}$ is a squared-error loss, epsilon is a hyperparameter, say, $\epsilon=0.2$. The truncated version of generalized advantage estimation is $\hat{A}_{k}$, $k$ specifies the time index in $[0, K]$, within a given length-$K$ trajectory segment, in which $\gamma$ is the discount factor determining the agent's horizon. $r_{k}(\theta)$ is the probability ratio, the behavior policy $\pi_{\theta}\left(a_{k} \mid s_{k}\right)$ denotes making an observation $s_{k}$ and taking an action $a_{k}$ at each timeslot $k$, $\theta$ is the parameter of actor neural network.   

\subsection{DRL Formulation of Examined Problem}
In this section, the large-scale V2G continuous charging/discharging coordination problem formulation by the PPO scheduling is presented as Fig.~\ref{fig:ppoflow}. When the EVs park at home, the DRL coordination strategy achieves peak shaving, valley filling and cost reduction by controlling the charging/discharging time and power of EVA. Besides, the proposed strategy is also used to reduce the volatility of renewable energy generation. 

We detail the DRL formulation of the examined large-scale V2G continuous charging/discharging coordination problem, the key elements of which are outlined in the following.

\paragraph{Agent} The examined EVA constitutes the agent, which gradually learns how to improve its retail charging/discharging decisions by utilizing experiences from its repeated interactions with the environment.

\paragraph{Environment} The environment consists of the EVs, power grid and RES (wind power and PV power), with all of which the EVA interacts.

\paragraph{State}

This initial state space $\mathcal{S}_i=(P^{k-23}, ..., P^k, SOC^k, {\sigma^{2}_k} )$ encapsulates three types of information:

(1) $(P^{k-23}, ..., P^k)$  denotes the past 24-hour power grid load values at the timeslot $k$; 

(2) $SOC^k$ represents the EVA SOC at the timeslot $k$;

(3) ${\sigma^{2}_k}$ indicates the load variance at the timeslot $k$.

In order to prevent gradient explosion of the neural network, the state values are normalized. The state at timeslot $k$ is defined as a vector $\mathcal{S}$. 
\begin{align}
\mathcal{S}(j)=\frac{\mathcal{S}_i(j)-Mean(\mathcal{S}_i)}{Std(\mathcal{S}_i)}.
\end{align}

Where $\mathcal{S}(j)$ represents the $j-th$ state value, which is normally distributed between 0 and 1. $Mean(\mathcal{S}_i)$ is the average of the set of initial state values, $Std(\mathcal{S}_i)$ is the standard deviation of the set of initial state values.

\paragraph{Action} 

Given the state $\mathcal{S}$, $a_{k}$ is the action of an agent at the timeslot $k$.
\begin{align}\label{action}
&\quad a_{k}=P_{EVA}^{k}.
\end{align}

The action $a_{k}$ represents the charging/discharging power of the EVA at the timeslot $k$. Let $a_{k}$ be positive when the EVA is charging and negative when discharging. Besides, we assume that the V2G equipment provides continuous charging/discharging power.

\paragraph{Reward}
Reward is the optimization goal for the agent and is the function of the load variance. To avoid sparse rewards during the agent learning process, the functions below highlight the rewards and penalties for every action undertaken by the agent while scheduling EVA charging/discharging.
\begin{align}
&r_{g}=\frac{\alpha}{\sigma^{2}}+\beta\left(P_{\max }-P_{\min }\right)+\chi\left\lvert  SOC_{\text {b}}^{k}-SOC^k\right\rvert,\label{rg}\\
&r_{r}={\frac{\alpha}{\sigma^{2}}+\beta\left(P_{\max }-P_{\min }\right)+\chi\left\lvert  SOC_{\text {b}}^{k}-SOC^k\right\rvert}+\frac\psi{f_1}.\label{rr}
\end{align}

Where $r_{g}$ is the reward of the agent in the baseload scenario, $r_{r}$ is the reward of the agent in the RES scenario.

\section{Experiment Results}

\subsection{Test System and Implementation}
\begin{figure}[h]
    \centering
    \includegraphics[width=0.45\textwidth]{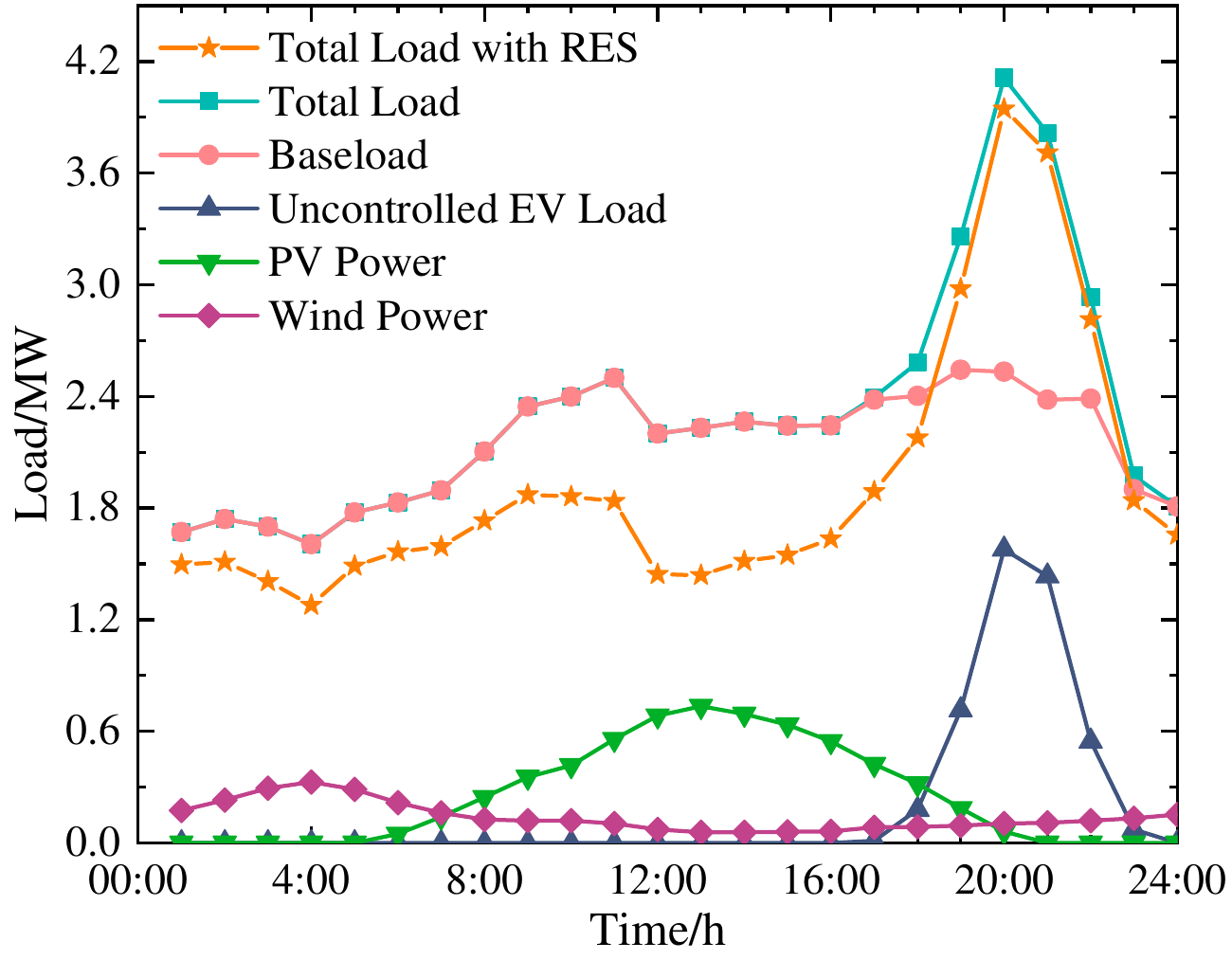}
    \caption{Daily load profiles with and without RES in a microgrid.}
    \label{fig:baseload}
\end{figure}

The following experiment results examine the EVA scheduling problem in the context of a single day with hourly resolution. The EVA serves 509 EVs for eliminating EV charging load peaks. The total load, the baseload and uncontrolled EV load in the examined day are shown as Figure~\ref{fig:baseload}, the total load profile represents the uncontrolled EV load plus the baseload profile. The objective function is set as (\ref{goal}). With RES, the total load of the microgrid results from the balance of EV, baseload, PV and wind power.

The assumed values of the remaining technical parameters of the EV are $p_{ch}^{max}= 6 kW$, $p_{dis}^{max}= -6 kW$, $Q_n=24 kWh$ for every EV. The detailed EV load model is shown in \cite{zhang2021optimal}. The arrival time is sampled from $N(18, 1^2)$ and is bounded between 15 and 21. For the departure time, its distribution $N(8, 1^2)$ is bounded between 6 and 10. The SOC of $n-th$ EV $SOC_n$ bounded between 0.2 and 0.8 is sampled from $N(0.5, 0.1^2)$. It should be noted that in the mentioned test systems in this article, the EVA is scheduled from timeslot 15 to 10. It is worth noting that the DRL coordination strategy does not rely on any knowledge of the distributions of these random variables. Thus, the proposed strategy can be transferred to different modeling mechanisms.  

\subsection{Training Process of PPO}
The PPO algorithm requires two networks, namely a critic network for evaluating a state and an actor network mapping a state into a probability distribution over the action space. Both networks are implemented with fully connected layers with the same input size. Both the critic and actor network have one output neuron and employ the rectified linearity units (ReLU) for all hidden layers. The Adam optimizer is employed for learning the neural network weights with a learning rate $lr^a=10^{-6}$ and $lr^c=2*10^{-6}$ for the actor and critic, respectively. For the critic, we use a discount factor of $\gamma = 0.95 $. We use a hyperparameter $\epsilon=0.2$ in order to avoid an excessively large policy update. We use $us = 10 $ as the critic and actor target network updating step. The output layer of the actor is a sigmoid layer to bound the continuous actions. We train with a mini batch size of $BS=32$ and for $M=3*10^{5}$ episodes, with $L=20$ hours per episode. Finally, we use $\alpha=10$, $\beta={-5}$, $\chi=10$ for the penalty weighting constant in (\ref{rg}).  

\begin{figure}[h]
\centering
\subfigure[Reward]{\includegraphics[width=0.24\textwidth]{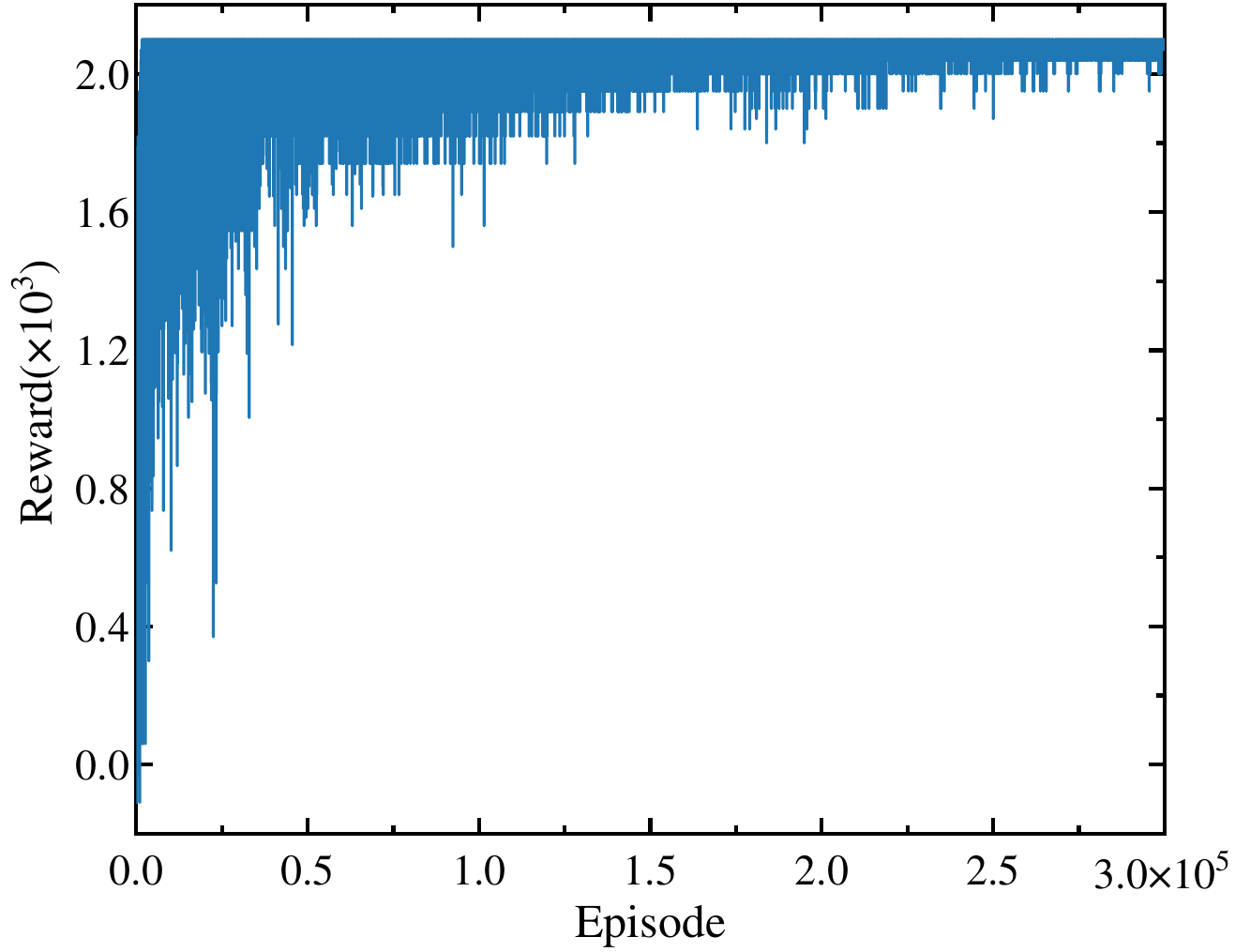}\label{reward}}
\subfigure[Load variance]{\includegraphics[width=0.24\textwidth]{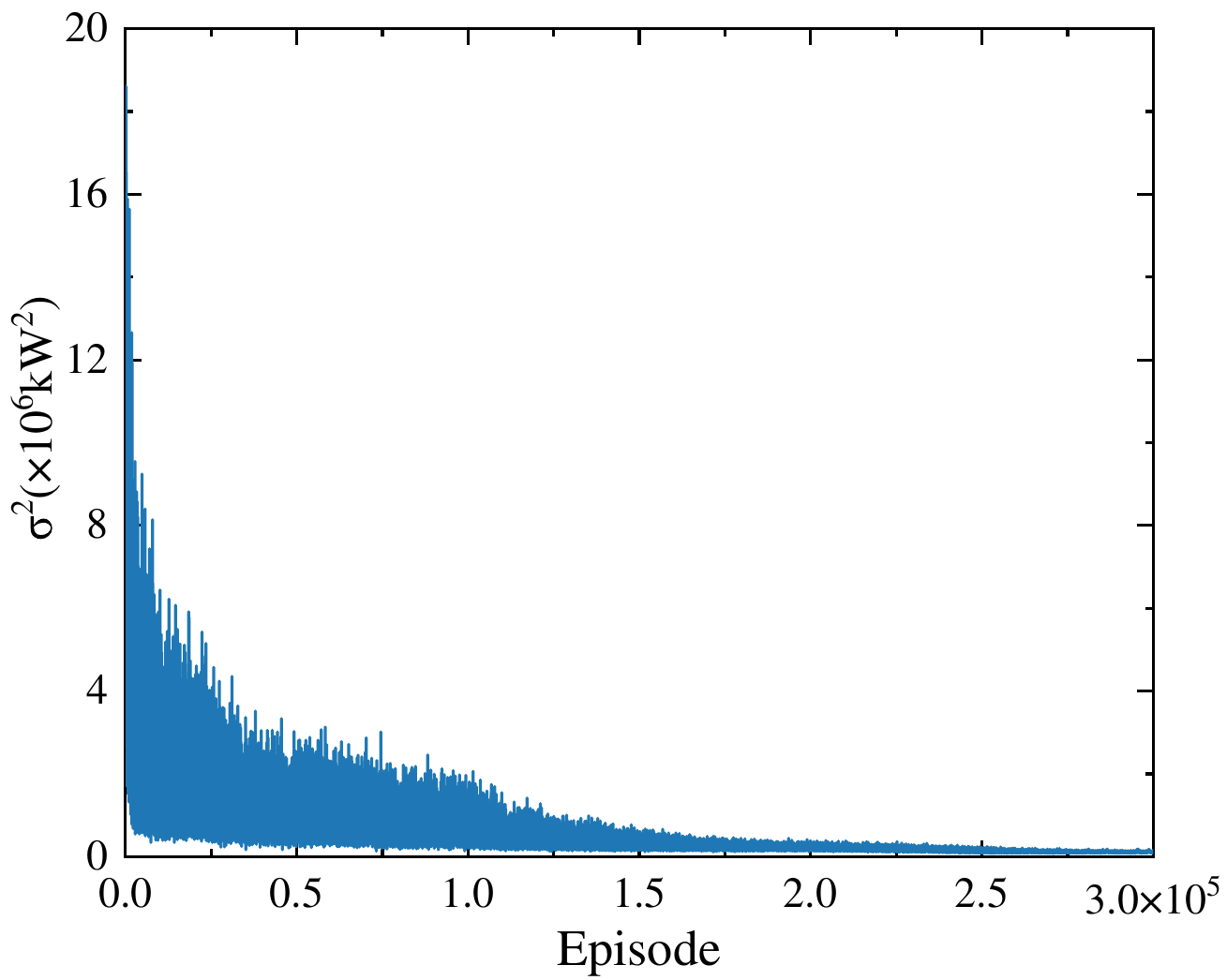}\label{loadva}}
\caption{Convergence process during the training.}
\label{fig:train}
\end{figure}

During the training process, the reward and the load variance over $3*10^{5}$ episodes are calculated and are illustrated in Fig.~\ref{fig:train}. 

As demonstrated in Fig.~\ref{fig:train}\subref{reward}, the charging/discharging schedule is randomly selected during the initial learning stages since the EVA is gathering more experiences by randomly exploring different, not necessarily profitable, actions. However, as the learning process progresses and more experiences are collected, the reward keeps increasing, and eventually converges around 2100 with small oscillations. This result demonstrates that the proposed strategy succeeds in learning a policy to maximize reward.

In Fig.~\ref{fig:train}\subref{loadva}, the load variance converges around 10663.4${kW}^2$ at episode $3*10^{5}$. Compared with the load variance of uncontrolled charging 419660.9${kW}^2$, the proposed strategy can reduce the load variance by 97.46$\%$.  After the training, the model optimized by the PPO algorithm is saved for transfer learning in different scheduling  scenarios.

\subsection{PPO Performance}
\begin{figure}[h]
\centering
\subfigure[Daily load profiles]{\includegraphics[width=0.24\textwidth]{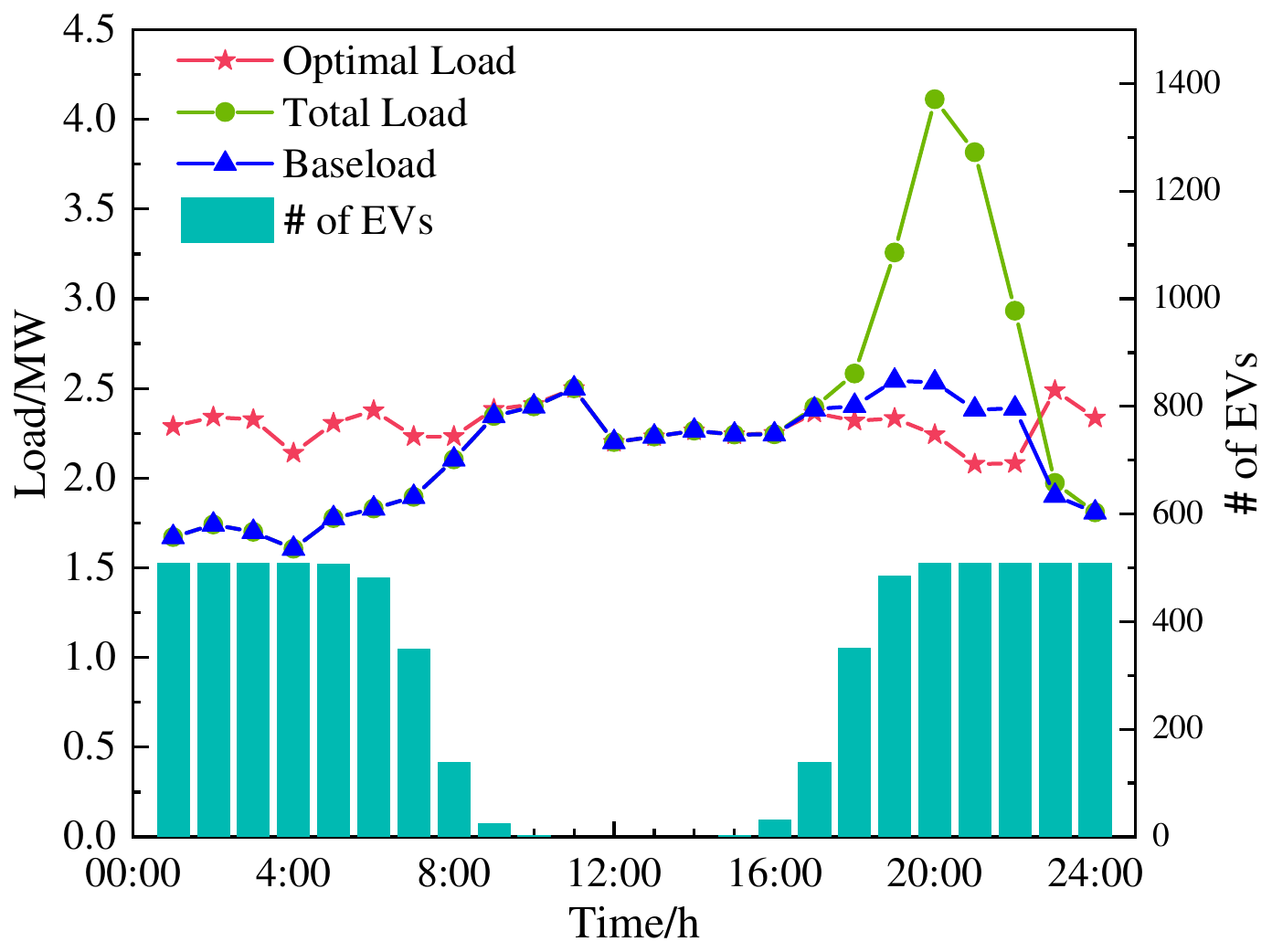}\label{loadpso}}
\subfigure[The load variance and action]{\includegraphics[width=0.24\textwidth]{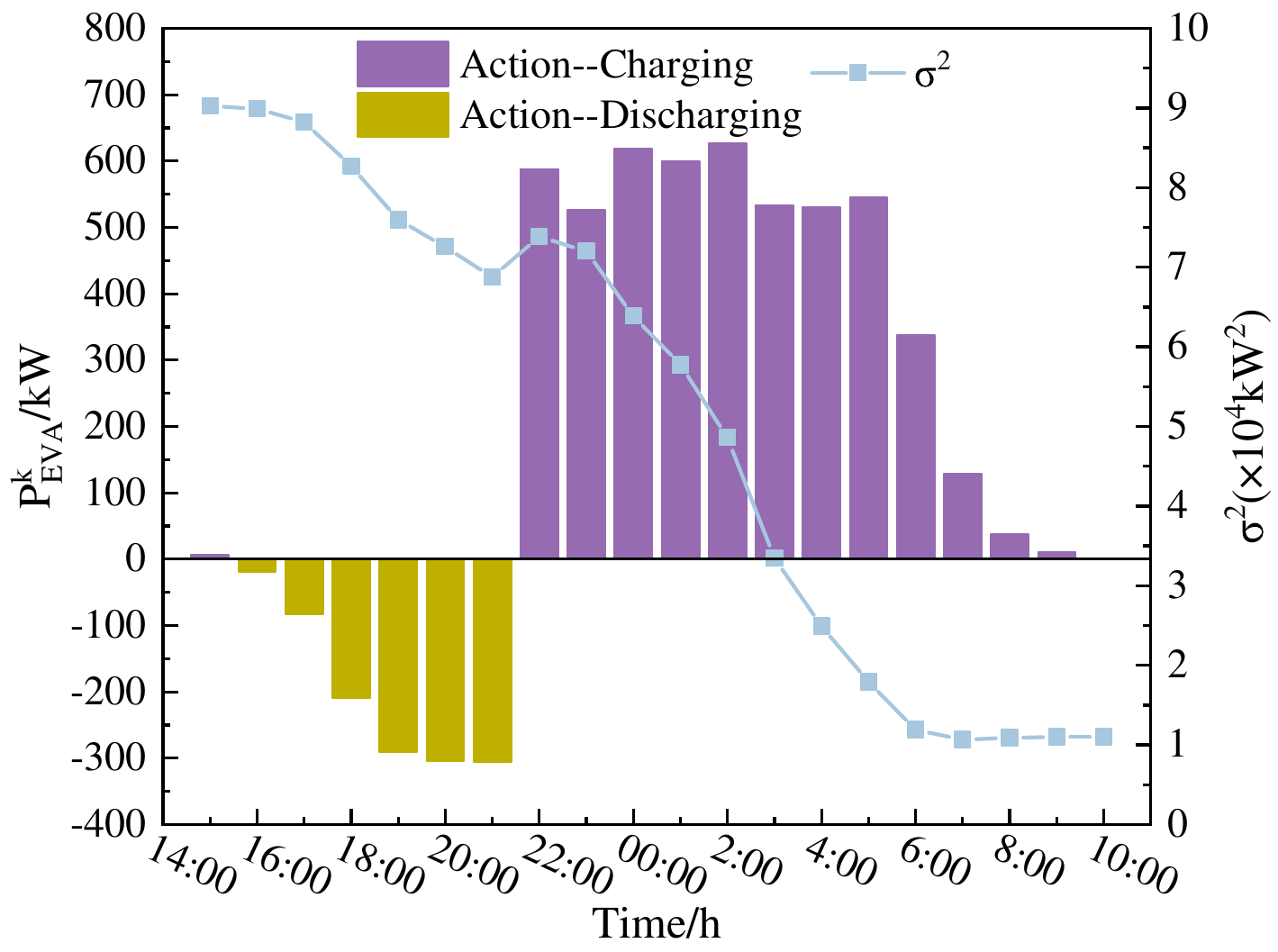}\label{variancea}}
\caption{Daily load profiles, load variance and action in the scheduling period.}
\label{fig:loadact}
\end{figure}

The performance of the proposed strategy is evaluated with the load variance reflecting the capability of the peak shaving and valley filling during scheduling operation. In a test day, the load variance is calculated as (\ref{variance}).  The load variance at each timeslot is calculated using the load profile of the previous 24 hours. 
In Fig.~\ref{fig:loadact}\subref{loadpso}, the optimal load profile is the total load profile with EVA scheduled according to the DRL coordination strategy and the blue bar represents the number of scheduled EVs in a day. The optimal load profile is flatter than the total load profile with uncontrolled EVA. The DRL coordination strategy by the PPO scheduling achieves peak shaving and valley filling by allocating the EVA charging demand to the load valley. 

At the arrival time of timeslot 15, the coordination scheduling starts to connect EVA to the power grid and the arrival load variance is 90269.2${kW}^2$. At the departure time of timeslot 10, the departure load variance becomes 11012.5${kW}^2$. During the scheduling period, the PPO agent gives charging/discharging actions according to the states of the load and the EVA SOC. EVA discharges at the peak timeslot 16-21 and is charged at the remaining time in the scheduling period. The load variance and charging/discharging power of EVA are shown in Fig.~\ref{fig:loadact}\subref{variancea}. 

\section{Transfer Learning Coordination}
The transfer learning ability of the DRL coordination strategy is demonstrated and evaluated with the three test systems, the RES test system, the large-scale EVA test system and the weekly test system.  
\subsection{Test Systems for Transfer}
In the RES test system, the RES are connected to the power grid as demonstrated in Figure~\ref{fig:baseload}, and we use $\alpha=10$, $\beta={-5}$, $\chi=10$, $\psi=-1$ for the penalty weighting constants in (\ref{rr}). The data of RES is shown in \cite{hou2018multiobjective}. In the large-scale EVA test system, EVAs with 5090, 20360, 35630, and 50900 EVs are considered where the baseload, the PV power and wind power combined is 100 times higher than the load value in the microgrid in Figure~\ref{fig:baseload}. 
In the weekly test system, the scheduling time is set as one week to evaluate the proposed strategy in long-time scale effect. In the weekly test system \cite{tuchnitz2021development}, uniformly distributed noises at ±10\% are introduced for the baseload and the RES. 

\subsection{EVA Scheduling with RES}
\begin{figure}[h]
\centering
\subfigure[Daily load profiles with RES integration]{\includegraphics[width=0.24\textwidth]{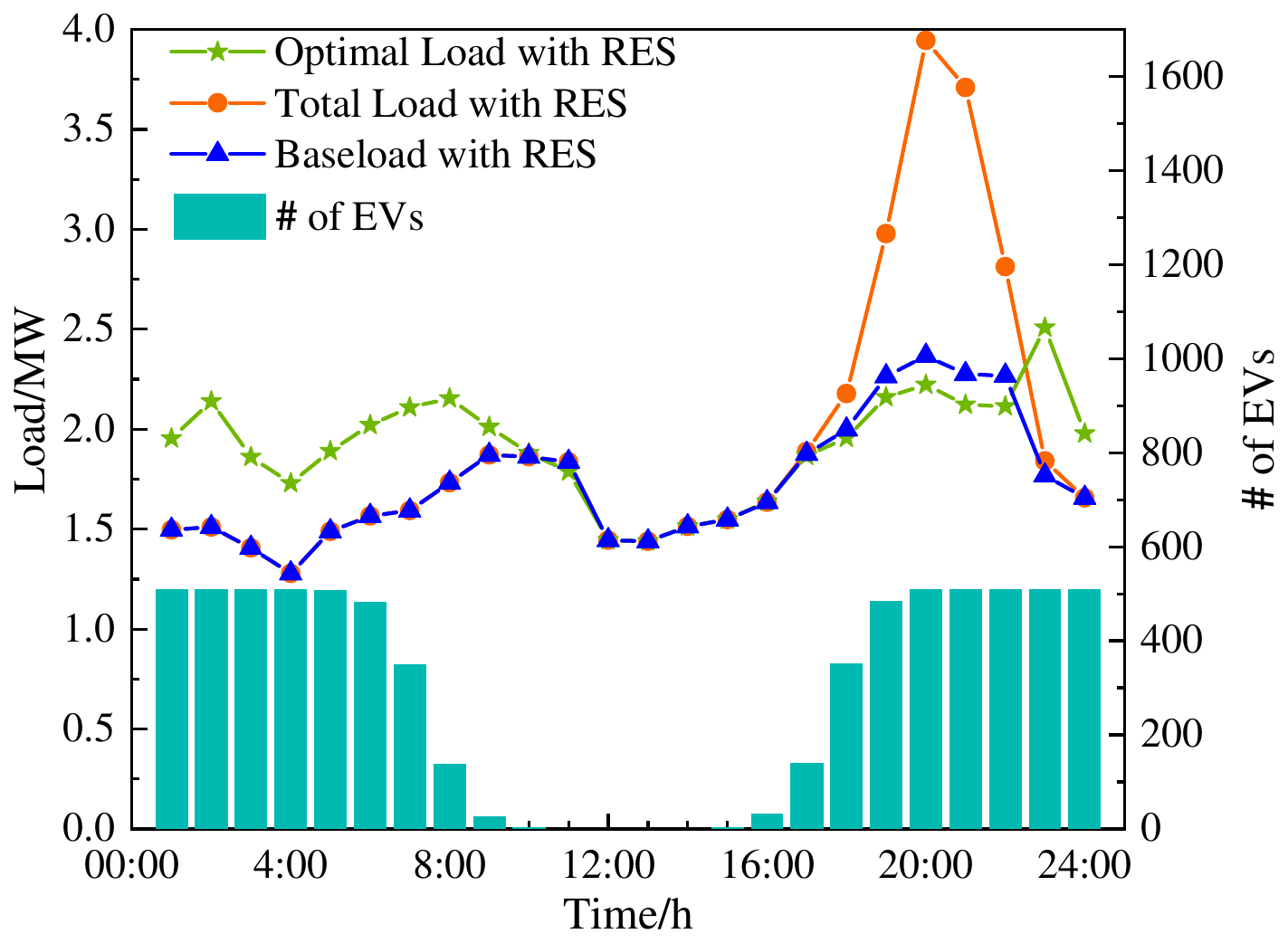}\label{loadpsopw}}
\subfigure[Daily optimal load profiles with different large-scale EVAs]{\includegraphics[width=0.23\textwidth]{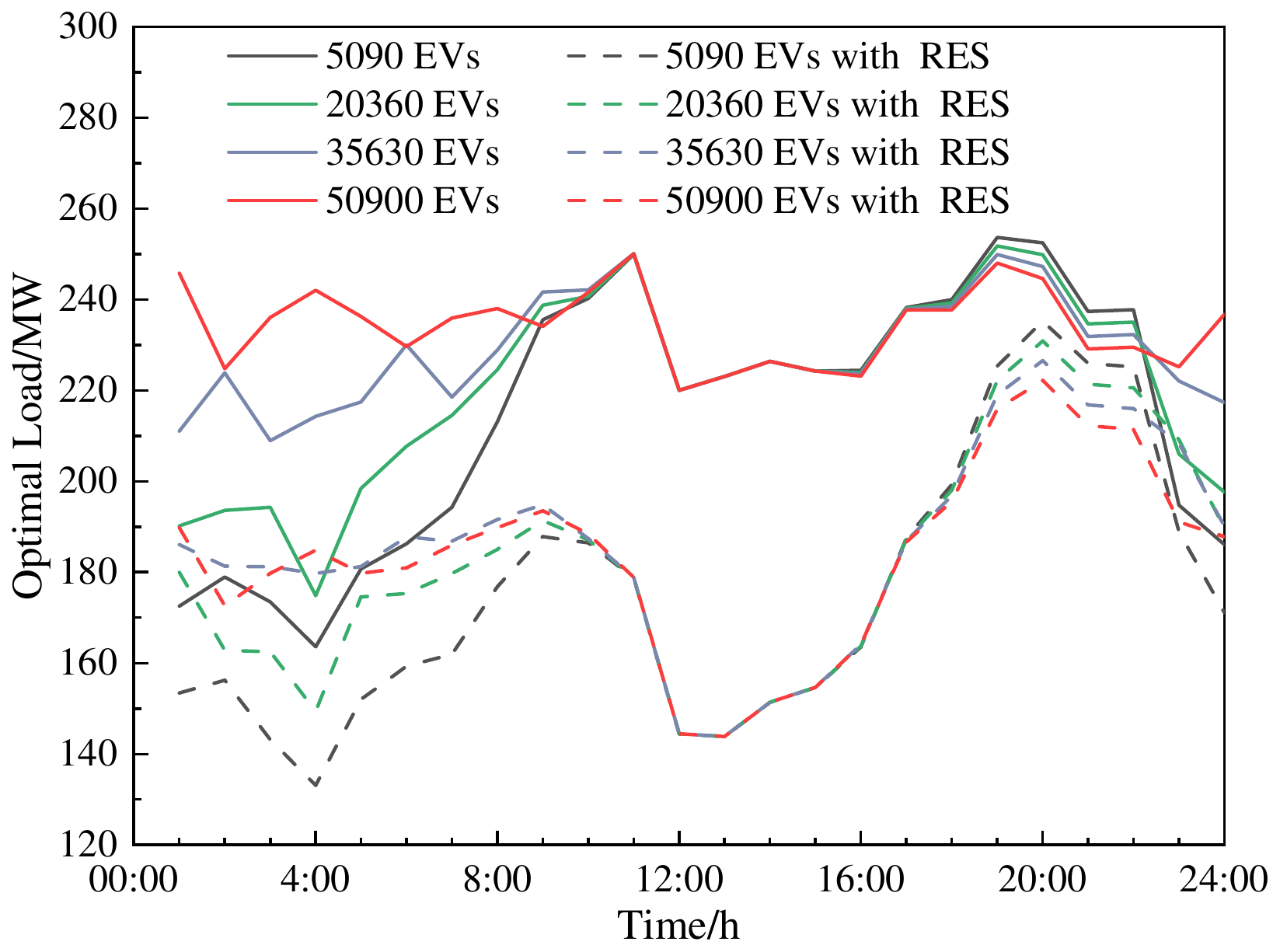}\label{loadlarge}}
\caption{Transfer learning daily load profiles with RES and large-scale EVA.}
\label{fig:load}
\end{figure}

The DRL agent gives charging/discharging actions according to the load values when  RES are connected in the RES test system. And the load variance decreases from 91688.4${kW}^2$ to 54127.0${kW}^2$. We can observe that the proposed strategy reduces the load variance by 88.89$\%$ in comparison with the uncontrolled charging.  With the PPO scheduling, the optimal load profiles with RES are shown in the figure~\ref{fig:load}\subref{loadpsopw}. Under the circumstance that the load fluctuation is intensified due to the RES, the proposed strategy still can effectively reduce the peak load and stabilize the power load with RES. 

\subsection{Large-scale EVA Scheduling}
The load profiles in the four large-scale EVA test systems with different numbers of EVs are demonstrated in Fig.~\ref{fig:load}\subref{loadlarge}. It can be seen that as EVA scales up, the effect of the peak shaving and valley filling become more prominent. Besides it, Fig.~\ref{fig:load}\subref{loadlarge} shows that the PPO scheduling can be transferred to the RES scenario directly with good performance as well. The optimal load variances for the four different EVAs are listed in Table~\ref{tab:pagenum8}. 
\begin{table}[h]
 \centering  
\caption{\textsc{Comparison of Optimal Load Variances for the Baseload and RES with Different EVA Sizes. }}
 \label{tab:pagenum8}
 \begin{tabular}{ccc}
 \toprule
 \makecell[c]{EVA ({\#} EV)}   &\makecell[c]{Baseload(${MW}^2)$}   &\makecell[c]{RES(${MW}^2)$}\\ 
   \midrule
  5090     &712.35$(34.61\% )^{\mathrm{a}}$ & 792.26$(33.59\%)$  \\
  20360    & 321.75 $(81.82\%)$& 569.44 $(73.34\%)$ \\
  35630    &137.27$(95.11\%)$   & 485.58$(85.86\%)$   \\
  50900    &70.97$(98.31\%)$ & 421.16  $(91.72\%)$  \\
  \bottomrule
\multicolumn{3}{p{230pt}}{$^{\mathrm{a}}$The percentage of the load variance reduction with respect to the uncontrolled charging are given in the bracket for the corresponding baseload and RES scenarios.}
 \end{tabular}
\end{table}
As shown in Table~\ref{tab:pagenum8}, the optimal load variance decreases with the increasing EVA size. Clearly, the larger EVA, the better DRL coordination strategy works in terms of the load variance performance. Besides it, the load variance for comparison is the departure time load variance. The scheduling optimization effect in the RES scenario is slightly worse than the baseload scenario, because the DRL coordination strategy doesn't work at noon when PV generates power.

\subsection{Weekly EVA Scheduling}
\begin{figure}[h]
    \centering
    \includegraphics[width=0.48\textwidth]{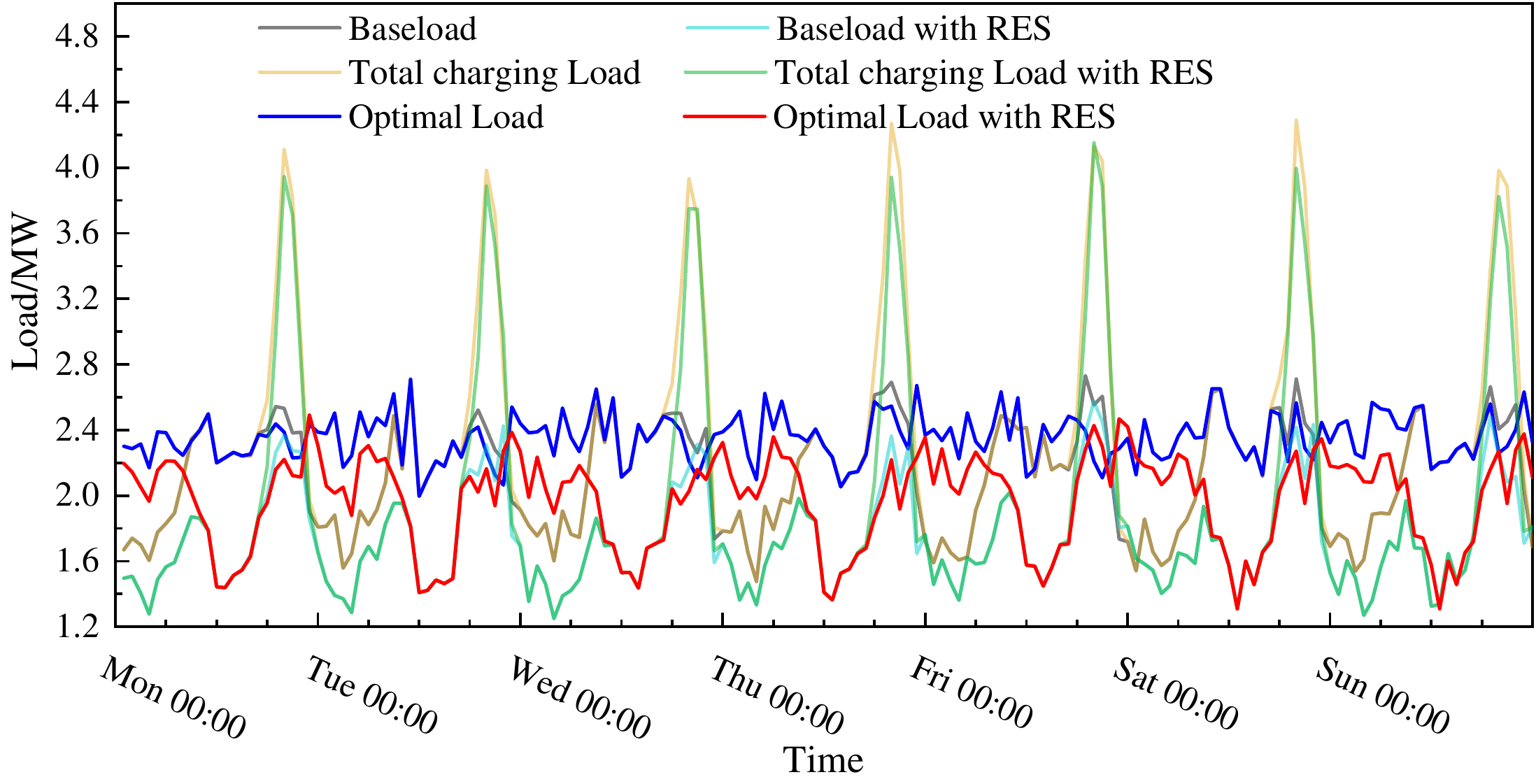}
		\caption{Charging/Discharging behavior in the weekly test system and optimal load profiles optimized by the PPO algorithm. }
		\label{fig:dispatchload1}
\end{figure}

The baseload profile, total load profile and optimal total profile in the weekly test systems with and without RES are shown in Fig.~\ref{fig:dispatchload1}. The DRL coordination strategy shows the strong advantages over the uncontrolled charging. The uncontrolled charging of EVs causes the large increasing of the existing peak loads and thus significantly causes the increasing of the load variance. Additionally, daytime load valleys cannot be scheduled, because the valleys are caused by RES and all EVs are not under scheduling. The green line represents the load profile of charging/discharging behavior generated by the PPO scheduling with perfect foresight. In the baseload scenario, the proposed strategy optimized by the PPO algorithm reduces the load variance by 95.04\% compared with the uncontrolled charging. Fig.~\ref{fig:dispatchload1} also shows the charging/discharging schedule with RES. The EVs are consistently charged at night during the load valleys and discharged during the load peaks, which promote the consumption and reduce the volatility of RES. In summary, the proposed strategy produces no not-enough-energy cases, no power-limit-exceeded steps, and reduces the load variance by 83.08\% compared with the uncontrolled charging with RES.

\subsection{Transfer Performance Comparison}
\paragraph{RES Test}
 The PSO algorithm\cite{arif2020analytical,gormez2021cost} and DQN algorithm \cite{mnih2015human,tuchnitz2021development} are widely used for the EV scheduling. The comparison between the baseload and RES scenarios is given in Table~\ref{tab:pagenum4}. 
\begin{table}[h]
 \centering
 \caption{\textsc{Comparison of Optimal Load Variances.} }
 \label{tab:pagenum4}
 \begin{tabular}{cccccc}
  \toprule
\makecell[c]{~} &\makecell[c]{PPO$_T$$^{\mathrm{b}}$} &\makecell[c]{PPO$_ R$$^{\mathrm{c}}$}  &\makecell[c]{PSO}   &\makecell[c]{DQN}     &\makecell[c]{Uncon$^{\mathrm{d}}$} \\ 
\midrule
\multirow{2}*{BL$^{\mathrm{a}}$}   & 11012.5  &10663.4     & 30910.0    &  33617.4 &419660.9 \\
 ~&$(97.37\%)$  &$(97.46\%)$   & $(92.63\%)$ &  $(91.99\%)$ &$(0)$  \\
\multirow{2}*{RES}   & 54127.0  &53440.4   & 71800.0    &  64279.5 &487292.5 \\ 
 ~&$(88.89\%)$  &$(89.03\%)$   & $(85.27\%)$ &  $(86.81\%)$ &$(0)$  \\
  \bottomrule
\multicolumn{6}{p{230pt}}{$^{\mathrm{a}}$The BL stands for Baseload,$^{\mathrm{b}}$ the PPO$_T$ stands for transfer PPO scheduling, $^{\mathrm{c}}$ the PPO$_R$ stands for retrained PPO scheduling, $^{\mathrm{d}}$ the Uncon stands for uncontrolled charging.}
 \end{tabular}
\end{table}

The PPO algorithm is directly transferred for the EVA scheduling. For comparison, PPO is retrained as well for the EVA scheduling. In the baseload scenario, the optimal load variances scheduled according to the retrained PPO and directly transfer PPO scheduling are 10663.4${kW}^2$ and 11012.5${kW}^2$, respectively. Compared with the load variance of uncontrolled charging 419660.9${kW}^2$, the the load variance can be reduced by 97.46$\%$ and 97.37$\%$, respectively. Besides it, the optimal load variances according to the PSO and DQN scheduling can be reduced by 92.63$\%$ and 91.99$\%$, respectively.
 
In the RES scenario, the retrained PPO and directly transfer PPO scheduling gives the load variances 53440.4${kW}^2$ and 54127.0${kW}^2$, respectively. The PSO and DQN scheduling can reduce the load variance by 89.03$\%$ and 88.89$\%$, respectively in reference with the uncontrolled charging 487292.5${kW}^2$. The PSO and DQN scheduling can reduce the load variance by 88.89$\%$ and 86.81$\%$, respectively.

For the comparison of the transfer learning performance, the difference of the peak shaving and valley filling effect between the retrained PPO and transfer PPO scheduling is not significant. The retraining process takes several hours and cannot be used for real-time scheduling of large-scale V2G. The above results and analysis can also prove the transfer learning effectiveness of our algorithm. What's more, after the training process, the DRL coordination strategy by the PPO scheduling can also be transferred to other scenarios. The iterative problem caused by excessive number of EVs is avoided. It takes about 1.26 ms to generate one schedule, which can meet the needs of real-time large-scale V2G scheduling.
As the PSO algorithm usually suffers from the curse of dimensionality due to the huge search space\cite{jafari2019using}, it cannot be transferred to the RES scenario, and the PSO scheduling results here are those obtained by retraining. 

\paragraph{Large-scale EVA Test}
The transfer ability of the proposed strategy has yielded good results in the PPO scheduling period of large-scale V2G. The optimal load variances according to the DQN scheduling are shown in Table~\ref{tab:pagenum7}. As the number of EVs increases, the load variance decreases. Besides, the effect of peak shaving and valley filling is very obvious in both scenarios. 
\begin{table}[h]
 \centering
 \caption{\textsc{Optimal Load Variances According the DQN Scheduling.}}
 \label{tab:pagenum7}
 \begin{tabular}{cccc}
 \toprule
 \makecell[c]{EVA({\#} EVs)}   &\makecell[c]{Baseload${MW}^2$)}   &\makecell[c]{RES(${MW}^2$)}\\ 
   \midrule
 5090     &794.02$(27.11\%)$ & 869.92$(27.08\%)$  \\
20360    & 519.28$(70.67\%)$ & 814.31$(61.88\%)$  \\
35630    &436.36 $(84.45\%)$  & 695.37$(79.75\%)$  \\
50900    &322.49$(92.31\%)$ & 652.72 $(87.17\%)$  \\
  \bottomrule
 \end{tabular}
\end{table}
As shown in Table~\ref{tab:pagenum8} and Table~\ref{tab:pagenum7}, the PPO scheduling is much better than the DQN scheduling in terms of peak shaving and valley filling effect with different large-scale EVAs. The superior performance of the PPO scheduling is driven by its ability to capture a continuous action space in contrast with the naive discretization approach adopted by DQN. 
The charging/discharging power of the EVA is determined by the number of EVs and their SOCs. When the scale of EVA changes in V2G, DQN needs to increase the discrete action dimension. DQN doesn't have good transfer ablity so that the DQN algorithm needs to be retrained and cannot be directly transferred to different scenarios. The PPO scheduling can adjust the charging/discharging power continuously, which makes the scheduling more accurate and efficient. 
The PPO scheduling performs significantly better in the peak shaving and valley filling effect than the existing DQN and PSO scheduling, and can meet the demand of real-time scheduling of large-scale V2G in microgrids with RES.

\paragraph{Weekly Test}
 The performance according to the PPO and DQN scheduling in a typical week are shown in Table~\ref{tab:pagenum6}. Compared with the uncontrolled charging in a typical week, the DQN scheduling can reduce the load variance by $88.90\%$. The PPO scheduling can reduce the load variance by 95.04\% much better than the DQN scheduling. In the microgrid with RES, the DQN scheduling\cite{tuchnitz2021development} can reduce the load variance by 81.85\%, the PPO scheduling can reduce the load variance by 85.96\% better than DQN.
\begin{table}[h]
 \centering
 \caption{\textsc{Comparison of Optimal Load Variances without and with RES.} }
 \label{tab:pagenum6}
 \begin{tabular}{cccc}
  \toprule
\makecell[c]{~} &\makecell[c]{PPO}     &\makecell[c]{DQN}     &\makecell[c]{Uncontrolled} \\ 
\midrule
\multirow{2}*{Baseload}   &20805.63        &  46551.97 &419264.32 \\ 
~& $(95.04\%)$    &  $(88.90\%)$ &$(0)$  \\
\multirow{2}*{RES}   &67073.08       &  86752.84 &477868.74 \\ 
~ &$(85.96\%)$    &  $(81.85\%)$ &$(0)$  \\

 \bottomrule
 \end{tabular}
\end{table}

\section{Discussion}
\subsection{Bounded SOC Constraints}
\begin{figure}[h]
\centering
\subfigure[Distribution of individual EV SOC]{\includegraphics[width=0.215\textwidth]{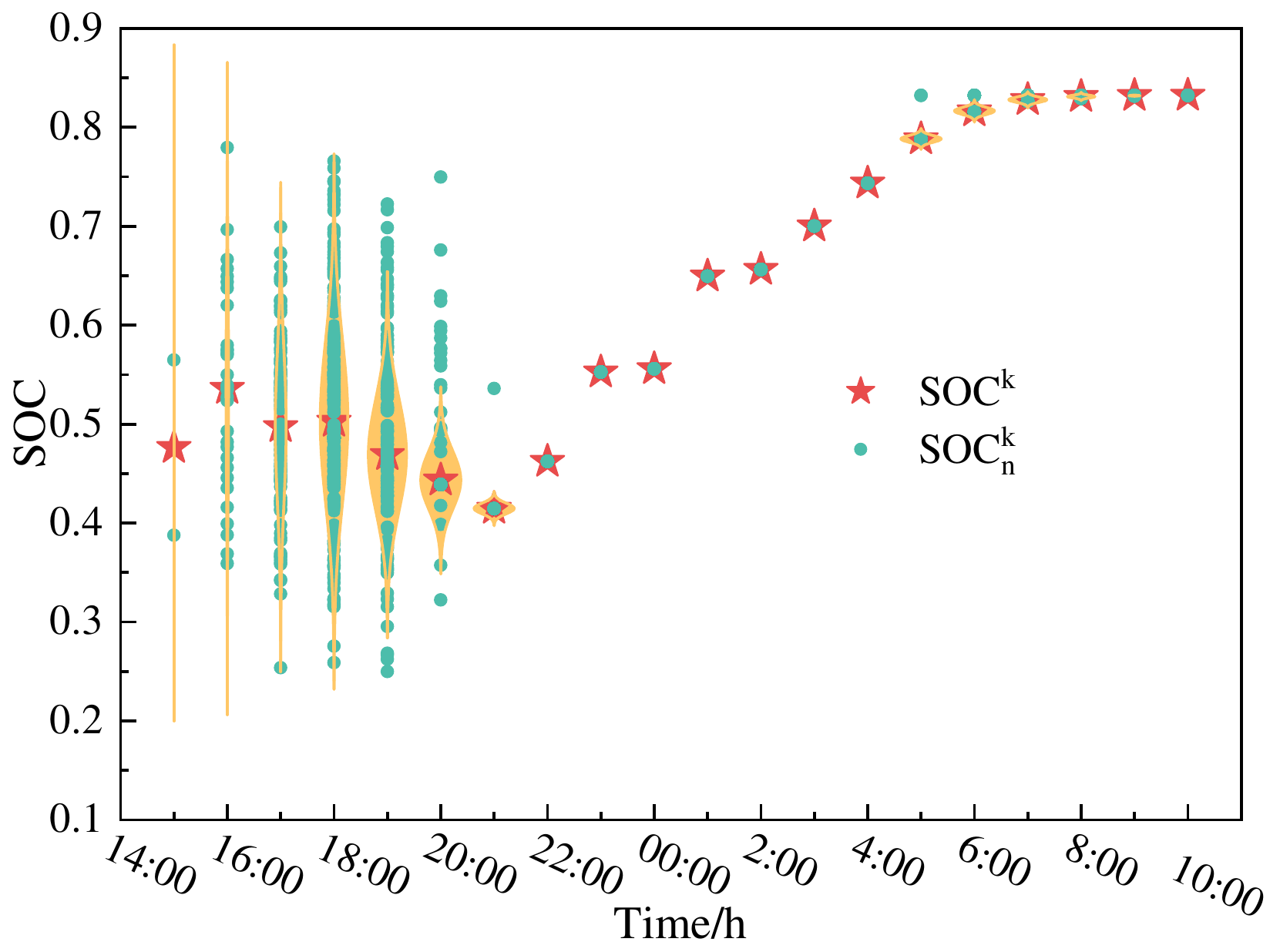}\label{soc509}}
\subfigure[Distribution of individual EV SOC with RES]{\includegraphics[width=0.215\textwidth]{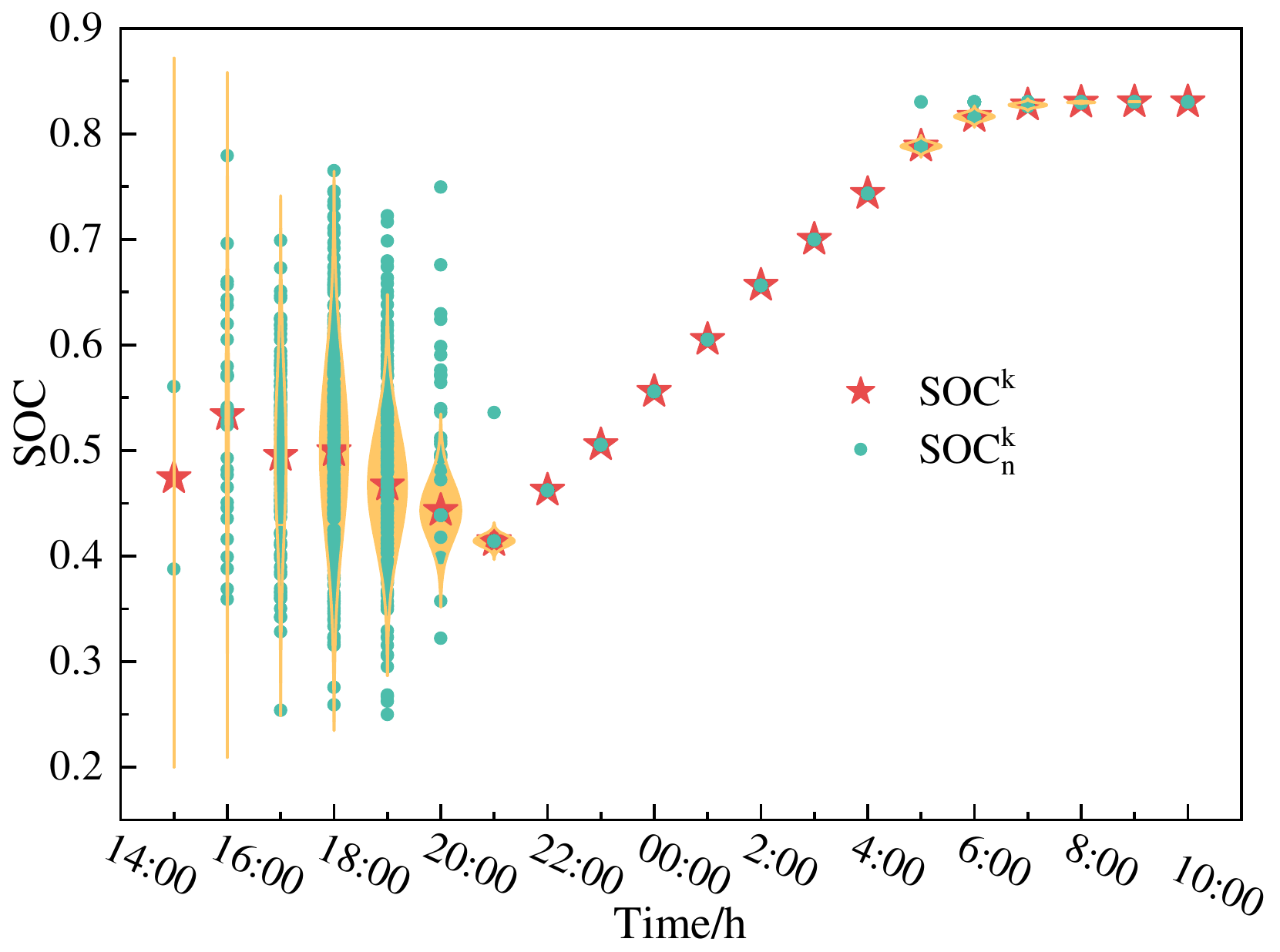}\label{soc509pw}}
\caption{Distribution of individual EV SOC with and without RES.}
\label{fig:soc}
\end{figure}

SOC is the key parameter to properly control the EV and to secure the power responses due to changes in operating conditions. The SOC constraints will ensure the DRL coordination strategy satisfy the driving demand of the EV users. The SOC constraints can reduce the long-term capacity fade rate and achieve higher number of equivalent full cycles or higher amount of cumulative discharge capacity over the battery's useful life.
\paragraph{Baseload}
 In Fig.~\ref{fig:soc}\subref{soc509}, the red star represents the EVA SOC $SOC^k$ in the scheduling period. The scheduling period is from timeslot 15 to 10. At departure time, the SOC of EVA is 0.83, which meets the departure SOC needs (Between 0.8 and 0.9). When the EVA scheduling is completed, the individual EV scheduling is required. The power allocation is elaborated in (\ref{kappa})-(\ref{soc1}). Therefore, the SOC of the $n-th$ EV all the time should satisfy (\ref{deps}). And the distribution of individual EV SOC  in the PPO scheduling period is shown in Fig.~\ref{fig:soc}\subref{soc509}. 
The EV SOCs in the scheduling time from timeslot 15 to 19 are more dispersed than in the remaining time, because EV SOCs differ a lot when EVs just arrive home for the charging/discharging scheduling. The scheduling is based on the EVA SOC. At the departure time, the SOC of individual EV is 0.83. In the scheduling period, the SOC of individual EV is always between 0.2 and 0.9, satisfying the SOC constraint of the individual EV. The bounded SOC constraints prevent battery overcharge and overdischarge, protect the battery from rapid degradation, and extend the service life of the battery. 
\paragraph{Transfer Learning Scheduling with RES}
The DRL coordination strategy has the transfer ability and can be used to schedule large-scale V2G continuous charging/discharging in microgrids with RES. In order to verify that the SOC of EVA in the transfer learning strategy does not cause to cross the boundary, the distribution of each EV SOC with RES is elaborated in Fig.~\ref{fig:soc}\subref{soc509pw}. At the departure time, the EVA SOC and the individual EV SOC are both 0.87, which also satisfies the SOC constraints.
\subsection{Cost Analysis}
In addition to the peak shaving and valley filling in terms of load variance reduction, the DRL coordination strategy also demonstrates the cost reduction benefit. Users can obtain the economic benefit from the discharging in the peak load shaving, because the TOU tariffs are set differently for the peak and valley hours of the grid load. For example, the TOU tariff is set at 0.8 CNY/kWh for shoulder hours, 0.4 CNY/kWh for off-peak hours, and 1.2 CNY/kWh for peak hours. The DRL coordination strategy reduces the EV charging cost by 76.56$\%$ in comparison to the uncontrolled charging with the charging cost of 9074.4 CNY. For the RES integration, the charging cost is reduced by 81.41$\%$.

\section{Conclusion}
This paper applies the  PPO algorithm in the DRL domain to large-scale V2G continuous charging/discharging coordination in microgrids with RES. The constrained EVA SOC and EVA power allocation are presented to schedule large-scale V2G efficiently. Then, taking the minimum load variance as the main objective, the constraint conditions such as RES and SOC of EVA and individual EV are considered. Finally, the PPO algorithm was adapted to optimize the EVA real-time charging/discharging power in continuous actions. Compared with swarm intelligence algorithms (e.g., PSO) and discrete DRL algorithms (e.g., DQN), the DRL coordination strategy optimized by the PPO is superior. Compared with uncontrolled charging, the DRL coordination strategy can reduce the load variance by 97.37$\%$ and the charging cost by 76.56$\%$ in the baseload test system, and reduce the load variance by 89.03$\%$ and the charging cost by 81.41$\%$ with RES. In a typical week, the proposed strategy can reduce the load variance by 95.04\% in the baseload scenario and reduce the load variance by 83.08\% with RES. Besides, the continuous DRL-based large-scale V2G coordinated charging/discharging strategy optimized by the PPO algorithm is independent of prediction information. Therefore, it can meet real-time optimization and EV SOC driving demands. In addition, large-scale V2G coordination has important reference significance for the orderly scheduling of energy storage systems in microgrids.  

Future work aims at incorporating the realistic variability of the state features (namely state of health and state of power) in the learning procedure of the DRL coordination strategy through the employment of data from large EV trials. This will help us test and enhance the generalization performance of the proposed strategy to render it robust against such variability, which constitutes a major current research challenge in the area of electric mobility.

\ifCLASSOPTIONcaptionsoff
  \newpage
\fi

\bibliographystyle{IEEEtran}
\bibliography{ref}

\begin{thebibliography}{10}
\providecommand{\url}[1]{#1}
\csname url@samestyle\endcsname
\providecommand{\newblock}{\relax}
\providecommand{\bibinfo}[2]{#2}
\providecommand{\BIBentrySTDinterwordspacing}{\spaceskip=0pt\relax}
\providecommand{\BIBentryALTinterwordstretchfactor}{4}
\providecommand{\BIBentryALTinterwordspacing}{\spaceskip=\fontdimen2\font plus
\BIBentryALTinterwordstretchfactor\fontdimen3\font minus
  \fontdimen4\font\relax}
\providecommand{\BIBforeignlanguage}[2]{{%
\expandafter\ifx\csname l@#1\endcsname\relax
\typeout{** WARNING: IEEEtran.bst: No hyphenation pattern has been}%
\typeout{** loaded for the language `#1'. Using the pattern for}%
\typeout{** the default language instead.}%
\else
\language=\csname l@#1\endcsname
\fi
#2}}
\providecommand{\BIBdecl}{\relax}
\BIBdecl

\bibitem{tookanlou2021optimal}
M.~B. Tookanlou, S.~A.~P. Kani, and M.~Marzband, ``An optimal day-ahead
  scheduling framework for e-mobility ecosystem operation with drivers’
  preferences,'' \emph{IEEE Transactions on Power Systems}, vol.~36, no.~6, pp.
  5245--5257, 2021.

\bibitem{singh2020cost}
J.~Singh and R.~Tiwari, ``Cost benefit analysis for v2g implementation of
  electric vehicles in distribution system,'' \emph{IEEE Transactions on
  Industry Applications}, vol.~56, no.~5, pp. 5963--5973, 2020.

\bibitem{dabbaghjamanesh2020stochastic}
M.~Dabbaghjamanesh, A.~Kavousi-Fard, and J.~Zhang, ``Stochastic modeling and
  integration of plug-in hybrid electric vehicles in reconfigurable microgrids
  with deep learning-based forecasting,'' \emph{IEEE Transactions on
  Intelligent Transportation Systems}, vol.~22, no.~7, pp. 4394--4403, 2020.

\bibitem{zavvos2019planning}
E.~Zavvos, ``Planning and analysing competing en-route charging stations for
  electric vehicles: A game-theoretic approach,'' Ph.D. dissertation,
  University of Southampton, 2019.

\bibitem{zhu2018coordinated}
X.~Zhu, M.~Xia, and H.-D. Chiang, ``Coordinated sectional droop charging
  control for ev aggregator enhancing frequency stability of microgrid with
  high penetration of renewable energy sources,'' \emph{Applied Energy}, vol.
  210, pp. 936--943, 2018.

\bibitem{li2021data}
C.~Li, Z.~Dong, G.~Chen, B.~Zhou, J.~Zhang, and X.~Yu, ``Data-driven planning
  of electric vehicle charging infrastructure: A case study of sydney,
  australia,'' \emph{IEEE Transactions on Smart Grid}, 2021.

\bibitem{peng2017optimal}
C.~Peng, J.~Zou, L.~Lian, and L.~Li, ``An optimal dispatching strategy for v2g
  aggregator participating in supplementary frequency regulation considering ev
  driving demand and aggregator’s benefits,'' \emph{Applied Energy}, vol.
  190, pp. 591--599, 2017.

\bibitem{hou2020multi}
H.~Hou, M.~Xue, Y.~Xu, Z.~Xiao, X.~Deng, T.~Xu, P.~Liu, and R.~Cui,
  ``Multi-objective economic dispatch of a microgrid considering electric
  vehicle and transferable load,'' \emph{Applied Energy}, vol. 262, p. 114489,
  2020.

\bibitem{jin2020optimal}
Y.~Jin, B.~Yu, M.~Seo, and S.~Han, ``Optimal aggregation design for massive v2g
  participation in energy market,'' \emph{IEEE Access}, vol.~8, pp.
  211\,794--211\,808, 2020.

\bibitem{mouli2017integrated}
G.~R.~C. Mouli, M.~Kefayati, R.~Baldick, and P.~Bauer, ``Integrated pv charging
  of ev fleet based on energy prices, v2g, and offer of reserves,'' \emph{IEEE
  Transactions on Smart Grid}, vol.~10, no.~2, pp. 1313--1325, 2017.

\bibitem{badawy2016power}
M.~O. Badawy and Y.~Sozer, ``Power flow management of a grid tied pv-battery
  system for electric vehicles charging,'' \emph{IEEE Transactions on Industry
  Applications}, vol.~53, no.~2, pp. 1347--1357, 2016.

\bibitem{mao2020electric}
H.~Mao, J.~Shi, Y.~Zhou, and G.~Zhang, ``The electric vehicle routing problem
  with time windows and multiple recharging options,'' \emph{IEEE Access},
  vol.~8, pp. 114\,864--114\,875, 2020.

\bibitem{li2019nash}
H.~Li, L.~Wang, D.~Lin, and X.~Zhang, ``A nash game model of multi-agent
  participation in renewable energy consumption and the solving method via
  transfer reinforcement learning,'' \emph{Proceedings of the CSEE}, vol.~39,
  no.~14, pp. 4135--4150, 2019.

\bibitem{lei2020deep}
L.~Lei, Y.~Tan, K.~Zheng, S.~Liu, K.~Zhang, and X.~Shen, ``Deep reinforcement
  learning for autonomous internet of things: Model, applications and
  challenges,'' \emph{IEEE Communications Surveys \& Tutorials}, vol.~22,
  no.~3, pp. 1722--1760, 2020.

\bibitem{sutton2011reinforcement}
R.~S. Sutton and A.~G. Barto, ``Reinforcement learning: An introduction,''
  2011.

\bibitem{huang2019adaptive}
Q.~Huang, R.~Huang, W.~Hao, J.~Tan, R.~Fan, and Z.~Huang, ``Adaptive power
  system emergency control using deep reinforcement learning,'' \emph{IEEE
  Transactions on Smart Grid}, vol.~11, no.~2, pp. 1171--1182, 2019.

\bibitem{yang2019two}
Q.~Yang, G.~Wang, A.~Sadeghi, G.~B. Giannakis, and J.~Sun, ``Two-timescale
  voltage control in distribution grids using deep reinforcement learning,''
  \emph{IEEE Transactions on Smart Grid}, vol.~11, no.~3, pp. 2313--2323, 2019.

\bibitem{wang2016dueling}
Z.~Wang, T.~Schaul, M.~Hessel, H.~Hasselt, M.~Lanctot, and N.~Freitas,
  ``Dueling network architectures for deep reinforcement learning,'' in
  \emph{International conference on machine learning}.\hskip 1em plus 0.5em
  minus 0.4em\relax PMLR, 2016, pp. 1995--2003.

\bibitem{wang2020deep}
B.~Wang, Y.~Li, W.~Ming, and S.~Wang, ``Deep reinforcement learning method for
  demand response management of interruptible load,'' \emph{IEEE Transactions
  on Smart Grid}, vol.~11, no.~4, pp. 3146--3155, 2020.

\bibitem{van2016deep}
H.~Van~Hasselt, A.~Guez, and D.~Silver, ``Deep reinforcement learning with
  double q-learning,'' in \emph{Proceedings of the AAAI Conference on
  Artificial Intelligence}, vol.~30, no.~1, 2016.

\bibitem{mnih2015human}
V.~Mnih, K.~Kavukcuoglu, D.~Silver, A.~A. Rusu, J.~Veness, M.~G. Bellemare,
  A.~Graves, M.~Riedmiller, A.~K. Fidjeland, G.~Ostrovski \emph{et~al.},
  ``Human-level control through deep reinforcement learning,'' \emph{nature},
  vol. 518, no. 7540, pp. 529--533, 2015.

\bibitem{lillicrap2015continuous}
T.~P. Lillicrap, J.~J. Hunt, A.~Pritzel, N.~Heess, T.~Erez, Y.~Tassa,
  D.~Silver, and D.~Wierstra, ``Continuous control with deep reinforcement
  learning,'' \emph{arXiv preprint arXiv:1509.02971}, 2015.

\bibitem{schulman2015trust}
J.~Schulman, S.~Levine, P.~Abbeel, M.~Jordan, and P.~Moritz, ``Trust region
  policy optimization,'' in \emph{International conference on machine
  learning}.\hskip 1em plus 0.5em minus 0.4em\relax PMLR, 2015, pp. 1889--1897.

\bibitem{schulman2017proximal}
J.~Schulman, F.~Wolski, P.~Dhariwal, A.~Radford, and O.~Klimov, ``Proximal
  policy optimization algorithms,'' \emph{arXiv preprint arXiv:1707.06347},
  2017.

\bibitem{chen2021electric}
L.~Chen, F.~Yang, S.~Wu, and Q.~Xing, ``Electric vehicle charging navigation
  strategy based on data driven and deep reinforcement learning,'' in
  \emph{Proceedings of the 5th International Conference on Control Engineering
  and Artificial Intelligence}, 2021, pp. 16--23.

\bibitem{wan2018model}
Z.~Wan, H.~Li, H.~He, and D.~Prokhorov, ``Model-free real-time ev charging
  scheduling based on deep reinforcement learning,'' \emph{IEEE Transactions on
  Smart Grid}, vol.~10, no.~5, pp. 5246--5257, 2018.

\bibitem{yan2021deep}
L.~Yan, X.~Chen, J.~Zhou, Y.~Chen, and J.~Wen, ``Deep reinforcement learning
  for continuous electric vehicles charging control with dynamic user
  behaviors,'' \emph{IEEE Transactions on Smart Grid}, vol.~12, no.~6, pp.
  5124--5134, 2021.

\bibitem{qian2022shadow}
T.~Qian, C.~Shao, X.~Wang, Q.~Zhou, and M.~Shahidehpour, ``Shadow-price drl: A
  framework for online scheduling of shared autonomous evs fleets,'' \emph{IEEE
  Transactions on Smart Grid}, 2022.

\bibitem{zhao2021dynamic}
Z.~Zhao and C.~K. Lee, ``Dynamic pricing for ev charging stations: A deep
  reinforcement learning approach,'' \emph{IEEE Transactions on Transportation
  Electrification}, vol.~8, no.~2, pp. 2456--2468, 2021.

\bibitem{qiu2020deep}
D.~Qiu, Y.~Ye, D.~Papadaskalopoulos, and G.~Strbac, ``A deep reinforcement
  learning method for pricing electric vehicles with discrete charging
  levels,'' \emph{IEEE Transactions on Industry Applications}, vol.~56, no.~5,
  pp. 5901--5912, 2020.

\bibitem{zhang2016virtual}
X.~Zhang, T.~Yu, B.~Yang, and L.~Li, ``Virtual generation tribe based robust
  collaborative consensus algorithm for dynamic generation command dispatch
  optimization of smart grid,'' \emph{Energy}, vol. 101, pp. 34--51, 2016.

\bibitem{kingma2014adam}
D.~P. Kingma and J.~Ba, ``Adam: A method for stochastic optimization,''
  \emph{arXiv preprint arXiv:1412.6980}, 2014.

\bibitem{zhang2021optimal}
Y.~Zhang, H.~Hou, J.~Huang, Q.~Zhang, A.~Tang, and S.~Zhu, ``An optimal subsidy
  scheduling strategy for electric vehicles in multi-energy systems,''
  \emph{Energy Reports}, vol.~7, pp. 44--49, 2021.

\bibitem{hou2018multiobjective}
H.~Hou, M.~Xue, Y.~Xu, J.~Tang, G.~Zhu, P.~Liu, and T.~Xu, ``Multiobjective
  joint economic dispatching of a microgrid with multiple distributed
  generation,'' \emph{Energies}, vol.~11, no.~12, p. 3264, 2018.

\bibitem{tuchnitz2021development}
F.~Tuchnitz, N.~Ebell, J.~Schlund, and M.~Pruckner, ``Development and
  evaluation of a smart charging strategy for an electric vehicle fleet based
  on reinforcement learning,'' \emph{Applied Energy}, vol. 285, p. 116382,
  2021.

\bibitem{arif2020analytical}
S.~M. Arif, A.~Hussain, T.~T. Lie, S.~M. Ahsan, and H.~A. Khan, ``Analytical
  hybrid particle swarm optimization algorithm for optimal siting and sizing of
  distributed generation in smart grid,'' \emph{Journal of Modern Power Systems
  and Clean Energy}, vol.~8, no.~6, pp. 1221--1230, 2020.

\bibitem{gormez2021cost}
M.~A. Gormez, M.~E. Haque, and Y.~Sozer, ``Cost optimization of an opportunity
  charging bus network,'' \emph{IEEE Transactions on Industry Applications},
  vol.~57, no.~3, pp. 2850--2858, 2021.

\bibitem{jafari2019using}
R.~Jafari, M.~M. Javidi, and M.~Kuchaki~Rafsanjani, ``Using deep reinforcement
  learning approach for solving the multiple sequence alignment problem,''
  \emph{SN Applied Sciences}, vol.~1, no.~6, pp. 1--12, 2019.

\end{thebibliography}

\end{document}